\documentclass[runningheads]{llncs}

\usepackage{eccv}

\usepackage{eccvabbrv}

\usepackage{graphicx}
\usepackage{booktabs}
\usepackage{multirow}
\usepackage{makecell}

\usepackage{pifont}
\usepackage[table]{xcolor}

\usepackage[accsupp]{axessibility}  

\usepackage{hyperref}

\usepackage{orcidlink}

\begin{document}

\title{MLVC: A Multi-platform Learned Video Codec for Real-World Deployment} 
\titlerunning{MLVC: Cross-Platform Neural Video Codec}

\author{Tanel Pärnamaa\orcidlink{0009-0005-4699-8958} \and
Martin Lumiste\orcidlink{0009-0003-5346-6840} \and
Ardi Loot\orcidlink{0000-0003-0251-119X} \and
Evgenii Indenbom\orcidlink{0009-0004-8765-6009} \and
Andrei Znobishchev\orcidlink{0000-0003-0412-2729} \and
Ando Saabas\orcidlink{0009-0001-9449-0404}}

\authorrunning{T.~Pärnamaa et al.}

\institute{Microsoft Corporation\\
\email{\{taparnam, ardiloot, eindenbom, aznobishchev, ansaaba\}@microsoft.com} \\
}

\maketitle

\begin{abstract}
Neural video codecs have surpassed classical codecs in coding efficiency but remain impractical for deployment due to cross-platform incompatibility and high computational cost.
Existing quantization-based solutions fail to produce deterministic results across diverse hardware platforms, leading to catastrophic decoding failures.
We introduce MLVC, a hardware-robust neural video codec designed for practical cross-platform inference.
The key idea is to explicitly transmit scale parameters through the hyperprior, which guarantees entropy coding consistency across devices without requiring bit-exact arithmetic.
While this increases bitrate overhead, we recover most of the coding efficiency through architectural improvements (gated memory, ReGLU activation), a long-term reference recovery mechanism, and domain-specific perceptual training.
On the VCD video conferencing benchmark, MLVC achieves \textgreater70\% BD-rate (MOS) improvement over hardware HEVC, the strongest deployable baseline, while reaching subjective quality competitive with DCVC-RT, which cannot operate across diverse platforms.
Both the encoder and decoder run at 100 FPS on average on commodity NPUs from Apple, Intel, and Qualcomm.
MLVC is the first neural video codec to combine competitive compression performance, real-time speed, and cross-platform robustness across diverse consumer devices, making it suitable for widespread deployment.
Code is available at \url{https://github.com/microsoft/mlvc}.
  \keywords{Neural Video Compression \and Cross-Platform Robustness \and Real-Time Video Processing}
\end{abstract}
\section{Introduction}
\label{sec:intro}

\begin{figure}[tb]
  \centering
  
  \includegraphics[width=\linewidth]{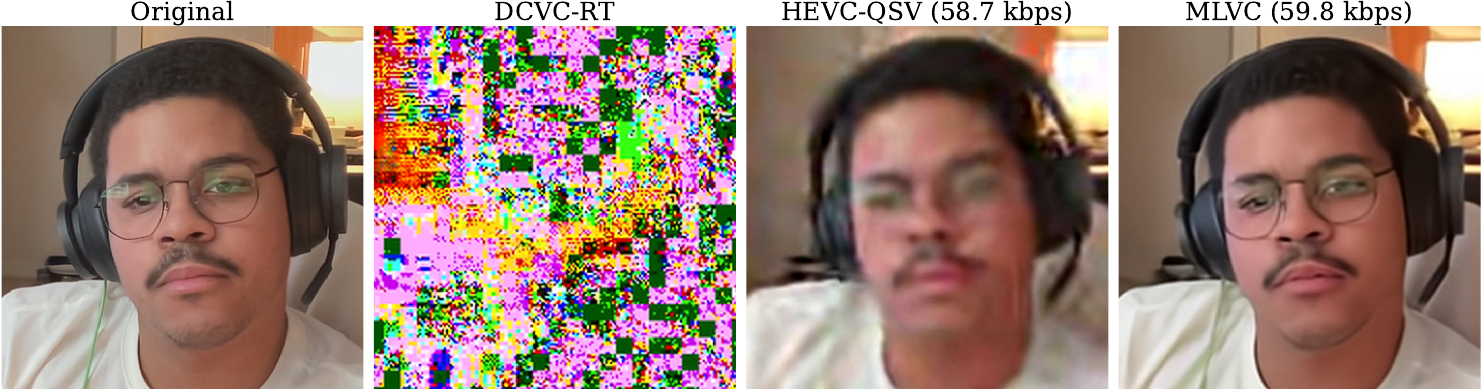}

  \caption{
  Neural codecs fail across diverse platforms.
  A video is encoded on an Apple M3 NPU and then decoded on another device (Intel Lunar Lake NPU).
  Existing state-of-the-art neural video codecs (e.g., DCVC-RT \cite{dcvc-rt}) suffer from catastrophic decoding failures due to numerical divergence between platforms, even when using quantization or calibration information \cite{tian2023towards}.
  Classical codecs like H.265 and the proposed neural codec MLVC both decode reliably across diverse platforms, while MLVC achieves significantly higher visual quality at the same bitrate.
   }
   \label{fig:frame_comparison_zoomed}
\end{figure}

Neural video compression has matured to the point where learned codecs consistently outperform traditional standards. \cite{mentzer2022neural, msra_video_gan, dcvc-fm, dcvc-rt}
Recent neural codecs achieve 60-70\% bitrate savings over H.265 at equivalent perceptual quality and surpass even the latest ECM standard in rate-distortion performance. \cite{dcvc-fm, dcvc-rt}
However, despite all this research progress, neural codecs remain unused in production systems like Zoom, Teams or WebRTC.

Two critical barriers prevent deployment.
The first is computational efficiency: most neural codecs are evaluated on high-end datacenter GPUs.
For practical impact in applications like video conferencing, neural codecs need to run in real-time on the neural processing units (NPUs) present in commodity consumer devices.

The second and more fundamental barrier is cross-platform compatibility.
Many applications, such as video conferencing systems, inherently require encoder and decoder to run on different devices, often with hardware from different vendors.
\Cref{fig:frame_comparison_zoomed} illustrates the severity of this problem: when we encode a video on an Apple M3 NPU and decode on an Intel NPU, the state-of-the-art neural codec DCVC-RT \cite{dcvc-rt}, even in its quantized form, produces completely corrupted output.

To address these barriers, we introduce MLVC, a neural video codec built for real world deployment. We specifically focus on model performance on NPUs rather than high-performance GPUs to show the practicality of our approach. Our contributions are as follows:
\begin{itemize}
\item We develop a scale sharing mechanism and transmit scale indices within the hyperprior, enabling reliable inference across diverse hardware platforms without requiring bit-exact arithmetic.
\item We introduce architectural improvements (gated memory for long-term modeling, hardware-compatible ReGLU activations), training and inference-time strategies (I-frame dropout, long-term reference recovery) that recover the coding efficiency lost to increased bitrate overhead, achieving \textgreater70\% BD-rate improvement over HEVC-QSV in subjective quality (MOS) on a video conferencing benchmark (VCD).
\item We demonstrate real-time performance (100+ FPS average for both encoding and decoding) on commodity NPUs from Apple, Intel, and Qualcomm, making MLVC the first neural video codec practical for widespread deployment on diverse consumer devices.
\end{itemize}

\section{Related Work}
\label{sec:related_work}

\subsection{Learned Compression}

The hyperprior architecture introduced by Ballé et al. \cite{balle2018variational} forms the foundation of modern learned compression \cite{cheng2020learned, he2022elic, dcvc-fm}.
This is commonly combined with autoregressive prior \cite{autoregressive-prior}.
Since fully autoregressive  priors are computationally prohibitive, recent work explores efficient approximations including checkerboard patterns \cite{checkerboard}, channel-wise autoregression \cite{minnen-charm} and hybrid spatial-channel models \cite{dcvc-hem}.
In learned video compression, numerous architectures since \cite{dvc} have advanced the field by improving motion modeling, temporal hierarchy, and entropy estimation \cite{scale-space-flow, HLVC, MLVC, qi2023motion, hu2021fvc, shi2022alphavc}. Video models often employ a third \textit{temporal} prior to exploit inter-frame correlation.
Li et al. \cite{DCVC} showed that contextual coding outperforms explicit residual coding, with subsequent advances often related to better context modelling \cite{dcvc-tcm, dcvc-dc}.
The current state-of-the-art models \cite{dcvc-fm, dcvc-rt} achieve improved rate-distortion performance over the reference software codec for ECM \cite{ECMref}. 

\subsection{Mobile and Low-Latency Codecs}

While most learned compression research prioritizes improving rate-distortion performance, a parallel direction focuses on computational efficiency for edge development. Codecs based on overfitting such as C3 \cite{c3} and Cool-chic video \cite{coolchic-video} achieve very low decoding complexities, but require costly encoding. MobileCodec \cite{mobilecodec} achieves 720p video decoding on a mobile phone with a neural accelerator. MobileNVC \cite{mobilevnc} extends this to 1080p video. DCVC-RT achieves over an 18$\times$ FPS speed-up compared to DCVC-FM \cite{dcvc-fm} on NVIDIA A100 GPUs, though edge device performance is not reported.

These works employ two main strategies for efficiency.
First, integer quantization to improve inference speed and also provide cross-platform consistency through bit-exact arithmetic.
However, cross-platform validation is limited: MobileNVC \cite{mobilevnc} reports no cross-platform testing, and DCVC-RT \cite{dcvc-rt} validates only across NVIDIA GPUs.
Since NVIDIA GPUs allow for low level kernel control, their validation demonstrates only single-vendor and not true cross-platform compatibility.
We will argue that these models are unlikely to work across heterogeneous NPU platforms.

Secondly, one of the main performance bottlenecks on edge devices has been the expensive pixel or feature space warping operator. As a result, most mobile methods redesign motion compensation to either use convolutions \cite{mobilecodec, dcvc-rt}, or simplify the warp operation \cite{mobilevnc}.

\subsection{Cross-platform Consistency}

The cross-platform consistency problem arises because entropy coding requires encoder and decoder to use identical probability distributions. In the case of DCVC-RT, the scale parameters $\boldsymbol{\sigma}$ characterize these distributions.
Even tiny numerical differences, for example those caused by non-deterministic floating-point computations, can cause the entropy decoder to select different lookup-table indices, resulting in incorrect latent values.
These errors cascade through temporal prediction, causing catastrophic failures (\cref{fig:frame_comparison_zoomed}).
We refer readers to Tian et al. \cite{tian2023towards} for detailed exposition.
Prior work addresses this through three main approaches, each with fundamental limitations.

\textbf{Network Quantization.} The most common approach quantizes entropy model 
computations to integer arithmetic, aiming for bit-exact results across platforms 
\cite{BalleJM19, dcvc-rt, he2022post, sun2021learned, device_interoperability_16bit, duan2023learned}.
In theory, the subnetwork responsible for producing $\boldsymbol{\sigma}$ could run in fixed-point integer arithmetic to guarantee cross-platform consistency, as traditional codecs do.
In practice, this is difficult to achieve on commodity NPUs.
First, many works adopt nonstandard bit-widths such as INT16 \cite{device_interoperability_16bit, quantized_decoder_16bit, dcvc-rt} and custom quantization scales that the NPU toolchains don’t support -- attempting to convert such models typically results in a compiler error, preventing acceleration entirely.
Second, even for the relatively standard INT8 convolution with INT32 accumulators, some NPUs (e.g. Apple NE) do not execute true INT8 but simulate it via FP16.\footnote{Vendor documentation indicates the int8-int8 compute path is only enabled starting from M4, released in 2024. \url{https://apple.github.io/coremltools/docs-guides/source/opt-overview.html}}
Third, across NPUs that do implement true INT8, compiler choices (kernel selection, operator fusion, reduction order), approximations, and rounding modes prevent bit-exact parity.
Unlike with traditional video codecs, there is no bit-exact reference specification for NPU vendors and quantized models can diverge more than analogous FP models due to compounding off-by-one errors. For these reasons, we use FP16 for inference, which is broadly supported across NPUs.
We present detailed cross-platform divergence measurements for both floating point and integer layers in the Supplementary Material.

\textbf{Calibration Information.} Tian et al. \cite{tian2023towards} transmit an additional side stream of \textit{calibration information} to decrease the likelihood of cross-platform index mismatches. 
While this reduces mismatch probability, it cannot provide hard guarantees - catastrophic failures remain possible, just less frequent. This probabilistic approach is insufficient for production deployment where reliability is critical.
Moreover, their experiments relied only on a single 96-frame video from the UVG dataset, encoded and decoded in FP32 precision.
Neural processing units typically use FP16 rather than FP32.
Since FP16 has a unit roundoff about four orders of magnitude larger, it is far more susceptible to rounding-error drift.
In our cross-device experiments (see supplementary material), the method fails in FP16: we observe float-index deviations $|\delta| > 0.5$, a situation where calibration cannot prevent index mismatches.

\textbf{Fixed-prior Approaches.}
More recently, Tian et al. \cite{tian2024effortless} proposed to transmit discrete codebook indices, eliminating entropy modelling entirely.
This effortlessly avoids catastrophic cross-platform failures.
However, their experimental results do not effectively validate competitive compression performance.
Most critically, their comparison uses mismatched GOP settings: GOP=32 for their model versus GOP=12 for H.264/H.265, giving their method an unfair advantage since classical codecs must encode expensive I-frames more frequently.

\section{Proposed Method}

\begin{figure}[tb]
  \centering
  \includegraphics[width=1.0\linewidth]{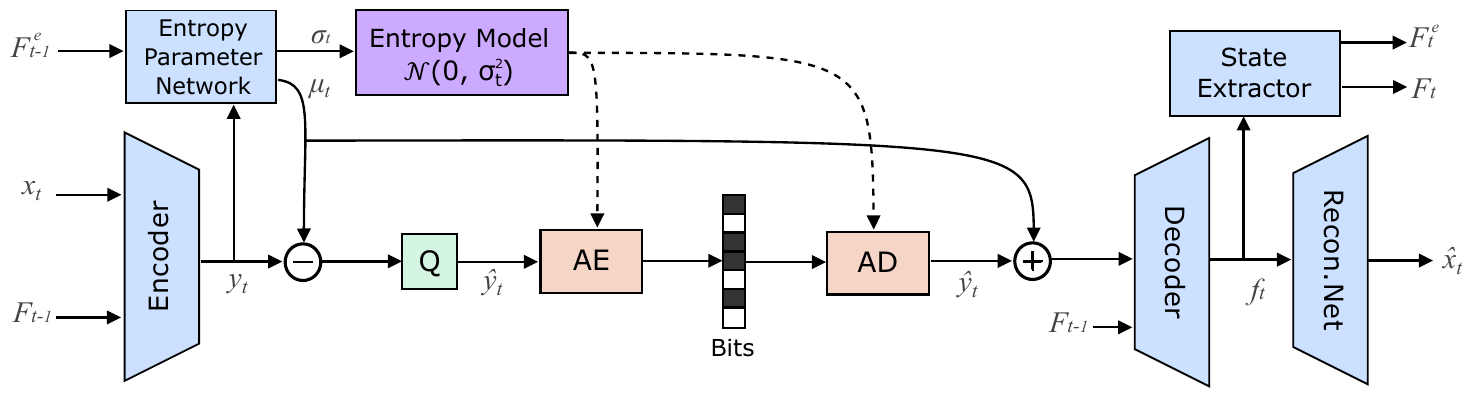}  
  \caption{Overview of the video coding framework adopted in this work, following modern learned codecs (e.g., DCVC-RT). The encoder maps the current frame $x_t$ and temporal context $F_{t-1}$ to the latent representation $y_t$. The Entropy Parameter Network, conditioned on $y_t$ and the temporal prior context $F_{t-1}^e$, produces the mean $\mu_t$ and scale $\sigma_t$ parameters. The mean is subtracted from $y_t$, and the residual is quantized (Q) and arithmetic encoded (AE) under a zero-mean Gaussian entropy model $\mathcal{N}(0,\sigma_t^2)$ to obtain the bitstream. On the decoder side, arithmetic decoding (AD) recovers $\hat{y}_t$, and the mean $\mu_t$ is added back. The decoder, further conditioned on temporal context $F_{t-1}$, produces the decoded feature $f_t$, from which the reconstruction network (Recon.~Net) yields the reconstructed frame $\hat{x}_t$. Finally, the state extractor derives the updated temporal contexts $F_t$ and $F_t^e$ from $f_t$ for the next time step.}
  \label{fig:architecture_overview}
\end{figure}

We use the DCVC-RT \cite{dcvc-rt} floating point model as a starting point, as it shows state-of-the-art compression ratio, while remaining lightweight and fast.
An overview of the video coding framework is shown in \cref{fig:architecture_overview}.
While our experiments focus on DCVC-RT, the proposed methods rely on components common to most modern video codec.
The scale-sending mechanism applies to any codec using adaptive entropy priors (e.g. hyperprior, autoregressive prior, temporal prior), the memory module to any recurrent codec maintaining temporal state, and the LTR mechanism to codecs that use reference features.

\subsection{Avoiding Catastrophic Failures}
\label{scale_sending_description}

Catastrophic decoding failures in existing neural video codecs are caused by mismatches in entropy model parameters between the encoder and decoder, specifically the scale parameters in DCVC-RT.
While eliminating all adaptive priors would trivially eliminate catastrophic decoding failures, this significantly degrades BD-rate performance in practice (see row \emph{0-prior} in Table \ref{tab:ablation}).
The alternative, transmitting all scale parameters explicitly, incurs prohibitive overhead, since there are $\frac{H}{16} \times \frac{W}{16} \times C_y$ scale parameters, comparable in magnitude to the encoded frame itself. 
However, scale parameters exhibit strong spatial and
cross-channel correlation that can be exploited for compression.
We reduce the parameter count by a factor
$s^2 \cdot r$ (typically 128×) through structured parameter sharing, then entropy-code the remaining parameters within the hyperlatent representation. This achieves entropy coding consistency with acceptable bitrate overhead.

\paragraph{Scale Sending Mechanism.}

We derive scale indices $\boldsymbol{I} \in \mathbb{N}_0^{C_y \times H_y \times W_y}$ from the quantized hyperlatent $\hat{\mathbf{z}} \in \mathbb{Z}^{C_z \times H_z \times W_z}$ through deterministic expansion.
Since $\hat{\mathbf{z}}$ is coded with a fixed factorized entropy model, it is identical on both the encoder and decoder sides, guaranteeing that the resulting scale parameters are also identical and thus avoiding catastrophic decoding failures. 
This design exploits two observations: (1) neighboring spatial locations often require similar scales, and (2) groups of channels often share statistical properties.

\begin{figure}[tb]
  \centering
  \includegraphics[width=1.0\linewidth]{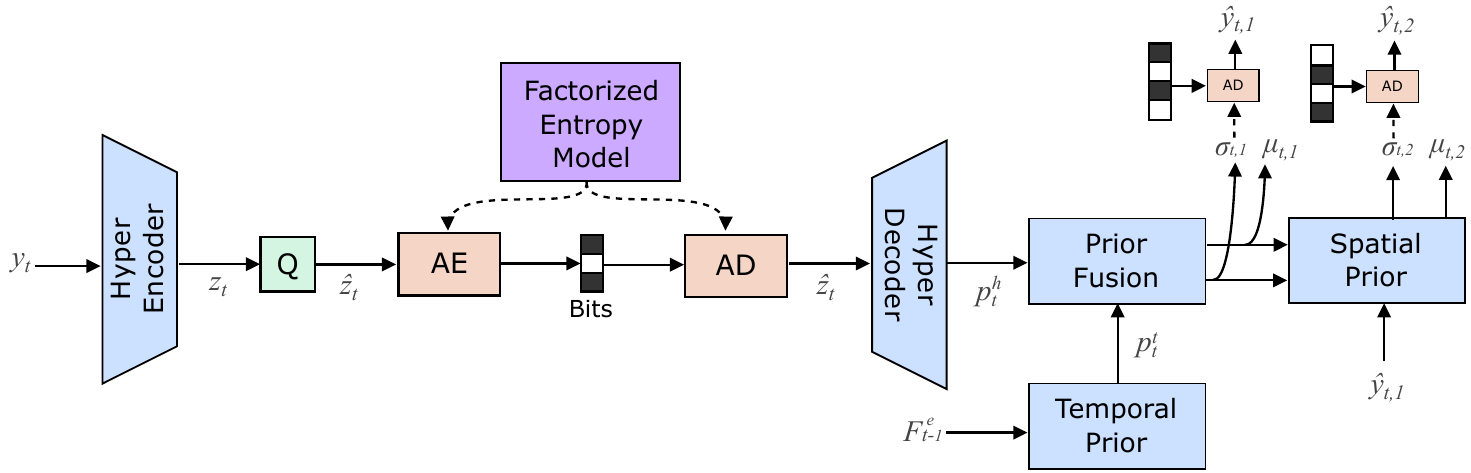}
  \caption{Entropy Parameter Network in modern learned codecs (e.g., DCVC-RT) incorporating hyper, temporal and spatial prior.
  The hyper encoder transforms $y_t$ into $z_t$, which is quantized and arithmetic coded using a factorized entropy model to yield $\hat{z}_t$. The hyper decoder then produces the hyperprior $p_t^h$.
  Because the factorized entropy model is identical and fixed on both the encoder and decoder sides, coding of $\hat{z}_t$ does not cause catastrophic decoding failures.
  The temporal prior maps the temporal context $F_{t-1}^e$ to $p_t^t$, and the Prior Fusion module merges $p_t^h$ and $p_t^t$ to produce the entropy parameters $(\sigma_{t,1}, \mu_{t,1})$ used to arithmetic code the first latent group $\hat{y}_{t,1}$.
  The Spatial Prior then conditions autoregressively on the already decoded $\hat{y}_{t,1}$ to produce $(\sigma_{t,2},\mu_{t,2})$ for the second group $\hat{y}_{t,2}$.
  However, the $\sigma_{t,i}$ can diverge between the encoding and decoding side, leading to catastrophic decoding failures.
  }
  \label{fig:standard_entropy_model_overview}
\end{figure}

\begin{figure}[tb]
  \centering
  \includegraphics[width=1.0\linewidth]{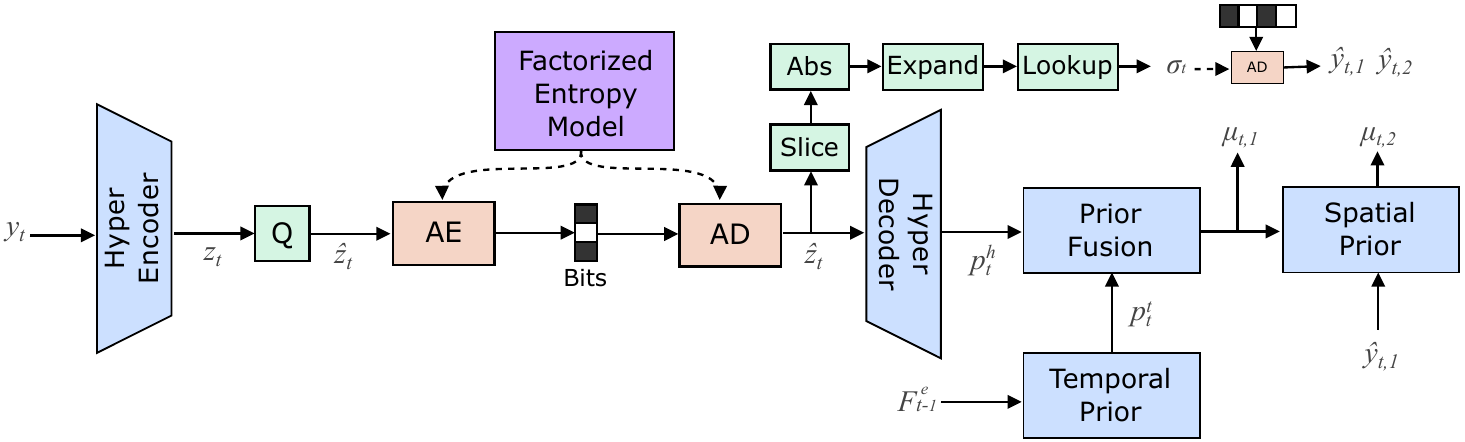}
  \caption{Entropy Parameter Network using the proposed scale-sending mechanism.
  A slice of the hyper decoder output is passed through Abs, Expand, and Lookup operations (detailed in \cref{fig:scale-sending-closeup}) to deterministically obtain the scale parameters $\sigma_t$ from the hyperlatent $\hat{z}_t$.
  Since $\hat{z}_t$ is coded with a fixed factorized entropy model, $\sigma_t$ is identical on both the encoder and decoder sides, avoiding catastrophic decoding failures.
  The mean parameters $\mu_{t,i}$ are still modeled autoregressively: Prior Fusion merges the hyper prior $p_t^h$ and temporal prior $p_t^t$ to produce $\mu_{t,1}$, and the Spatial Prior conditions on the already-decoded $\hat{y}_{t,1}$ to produce $\mu_{t,2}$.
  A further benefit over the standard approach (\cref{fig:standard_entropy_model_overview}) is decoding speed: because $\sigma_t$ is available before decoding any latent group, all groups can be arithmetic-decoded in a single AD pass. In contrast, the standard approach requires interleaving arithmetic decoding and neural network calls for each group.
  }
  \label{fig:scale_sending_entropy_model_overview}
\end{figure}

\begin{figure}[tb]
  \centering
  \begin{subfigure}[t]{0.52\linewidth}
    \centering
    \includegraphics[width=\linewidth]{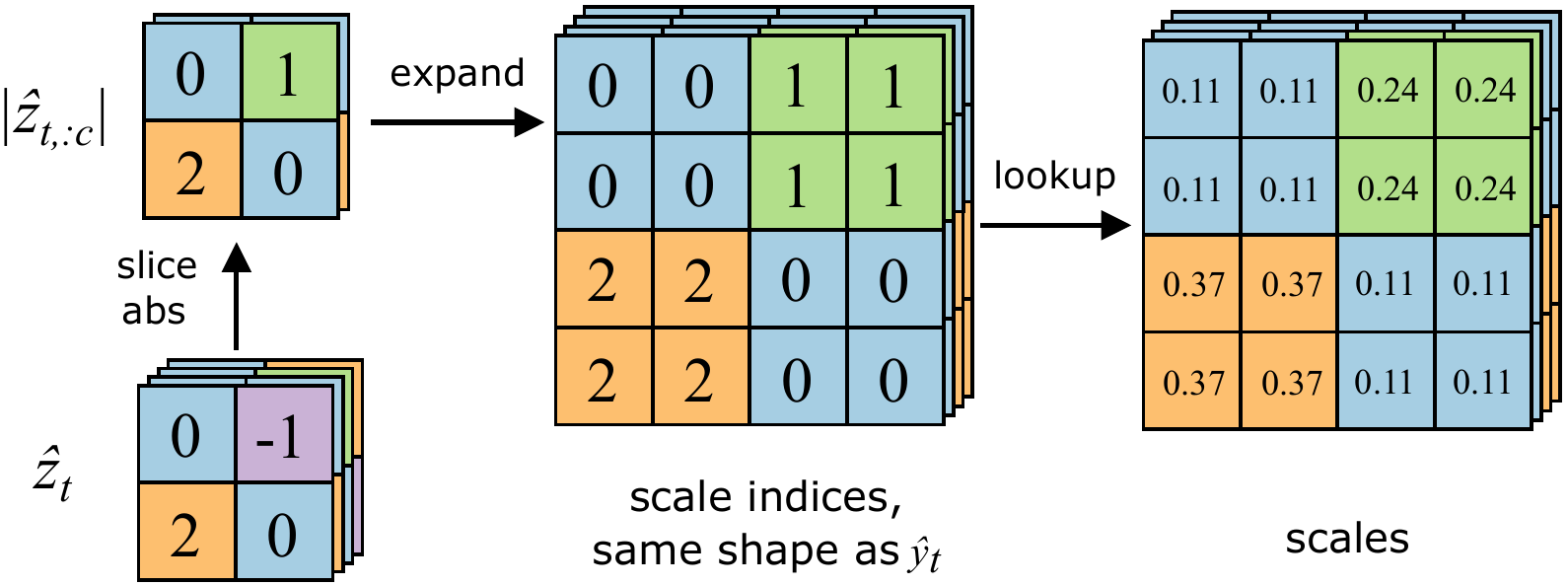}
    \caption{Scale sending architecture.}
    \label{fig:scale-sending-closeup}
  \end{subfigure}\hfill
  \begin{subfigure}[t]{0.42\linewidth}
    \centering
    \includegraphics[width=\linewidth]{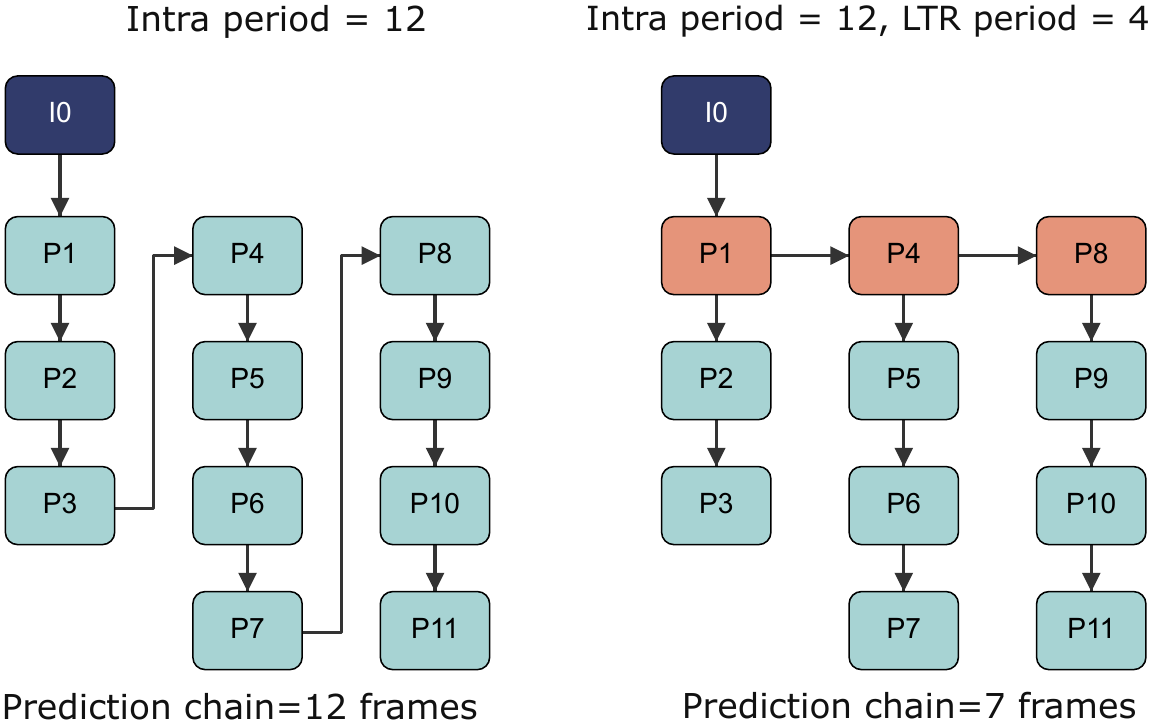}
    \caption{LTR prediction structure.}
    \label{fig:long-term-reference}
  \end{subfigure}

  \caption{(a) A visualization of the steps of the scale-sending mechanism described in \cref{scale_sending_description} (b) Comparison of prediction structures with and without long-term reference (LTR) frames. (b, left) Standard encoding with I-frame period of 12 creates prediction chains up to 12 frames long. (b, right) Using LTR period of 4 (starting from frame 1) reduces maximum chain length to 7 frames and improves robustness to packet loss. In practice, LTR period is another inference parameter, similar to intra and reset period.}
  \label{fig:scale_sending_overview}
\end{figure}

Let $r \in \mathbb{N}$ denote the channel repetition factor and $s = H_y/H_z$ the spatial expansion factor. The generation process comprises three steps:

\begin{align}
\boldsymbol{I}^{\text{base}} &= |\hat{\mathbf{z}}_{1:C_y/r}| \in \mathbb{N}_0^{C_y/r \times H_z \times W_z} \\
\boldsymbol{I}_{c,h,w} &= \boldsymbol{I}^{\text{base}}_{\lfloor c/r \rfloor,\, \lfloor h/s \rfloor,\, \lfloor w/s \rfloor} \\
\boldsymbol{\sigma}_{c,h,w} &= \operatorname{lookup}(\boldsymbol{I}_{c,h,w})
\end{align}
where $c \in [0, C_y)$, $h \in [0, H_y)$, $w \in [0, W_y)$. The absolute value ensures non-negative indices, the expansion maps each hyperlatent position to an $s \times s$ spatial block and replicates each channel $r$ times, and the final lookup retrieves quantized scale values from a fixed codebook.
Notably, the same scale parameters are shared across all groups in the autoregressive prior, so all groups can be arithmetic-decoded in a single pass without interleaving arithmetic decoding and neural network calls.
\Cref{fig:standard_entropy_model_overview} shows the standard entropy model of DCVC-RT, \cref{fig:scale_sending_entropy_model_overview} shows the proposed scale-sending variant, and \cref{fig:scale-sending-closeup} details the individual steps.

\subsection{Reducing Divergence}
\label{sec:reducing-divergence}

While our scale-sending approach eliminates catastrophic entropy decoding failures, floating-point differences still cause gradual divergence in the features and reconstructed frames.
Since encoder and decoder maintain separate feature  buffers that are updated recursively, even small per-frame differences compound over long prediction chains.
Without a synchronization mechanism, the divergence can grow arbitrarily large.
We address this through three strategies: a novel prediction chain control via long-term reference recovery inspired by traditional codecs, hardware-compatible activation functions, and periodic synchronization via I-frames.

\paragraph{Long-Term Reference Recovery.}
The most straightforward method for mitigating divergence over long prediction chains is to reduce the chain length by employing a small I-frame period (e.g., 64 frames). Although this approach can effectively address divergence, it incurs a substantial cost in terms of rate-distortion performance due to the use of expensive intra frames. As an alternative, a less costly mechanism is proposed, inspired by long-term reference (LTR) recovery techniques in traditional codecs. Figure \ref{fig:long-term-reference} illustrates the core concept of this approach: rather than relying exclusively on I-frames, proactive LTR recovery frames are periodically introduced to manage the prediction chain. This strategy reduces the frequency of expensive I-frames while maintaining the same number of frames in the prediction chain, resulting in an overall improvement in BD-rate.

Beyond reducing the prediction chain, proactive LTR recovery also preserves its original advantage of enhancing codec robustness against packet loss by providing periodic recovery frames without explicit requests, as in traditional codecs.
It should be noted that, on a single platform with a fixed intra period, the LTR mechanism would strictly regress BD-rate due to the introduction of additional redundancy. However, in cross-platform scenarios, this reduction in prediction chain length can improve compression efficiency, as reference divergences may otherwise lead to significant quality degradation in the absence of LTR.

\paragraph{Hardware-Compatible Activations.}
Many NPUs implement nonlinear activation functions through piecewise approximations rather than exact computation. Since these approximations are vendor-specific and unstandardized, even small differences compound through the deep network, causing cross-platform divergence. Furthermore, complex activations such as WSiLU often lack optimized NPU kernels, resulting in reduced inference speed. We perform a sweep over common activation functions supported by Apple Neural Engine \cite{coreml-ops} and find that only ReLU and LeakyReLU incur no error compared to the true value, we therefore restrict our architecture to use these. While simpler activations incur a BD-rate penalty, they are necessary for reliable cross-platform operation.

\paragraph{Regular I-Frame Period.}
Even with LTR recovery and simple activations, feature buffers still diverge over extended sequences due to fundamental floating-point precision limits.
We therefore employ periodic I-frames to fully resynchronize encoder and decoder feature states. 
While our model is able to handle very long prediction chains on a single platform, regular I-frames are necessary for practical deployment across multiple hardware platforms.
This periodic synchronization incurs BD-rate overhead compared to single-platform operation but remains necessary for reliable cross-platform decoding.
This differs from traditional codecs, which could run cross-platform without periodic I-frames.

\subsection{Unified I-Frame and P-Frame Model}

Most learned video codecs employ separate models for I-frames and P-frames, even though these architectures share substantial structural overlap.
Maintaining two distinct models increases both storage requirements and deployment complexity.
For instance, in DCVC-RT \cite{dcvc-rt}, the I-frame and P-frame models require 174 MB and 79 MB respectively in FP32.
To reduce this overhead, we simplify to a single unified model by defining I-frames as P-frames with a uniform gray reference image (0.5 in YUV colorspace).
This eliminates the need for a separate I-frame model while maintaining the ability to encode frames without temporal dependencies.
Although this simplification introduces a BD-rate degradation, it reduces engineering and deployment complexity.
Given that practical systems prioritize simplicity and maintainability, this trade-off is well justified.

\subsection{Improving Quality}
\paragraph{Memory.}
While DCVC-RT employs a recurrent architecture with features serving as temporal state, this representation has limited capacity for preserving long-term information, which is important for reconstructing temporarily occluded objects or maintaining consistency across frames.

Drawing inspiration from gated recurrent architectures \cite{lstm}, we augment the decoder with an explicit long-term memory state that enables the model to selectively retain and retrieve information over longer temporal horizons. The memory-enhanced decoder operates as follows:

\begin{equation}
\begin{aligned}
\mathbf{f}_{in} &= \text{Decoder}(\hat{\mathbf{y}_t}, \mathbf{F}_{t-1}) \\
[\mathbf{s}_t, \mathbf{g}_f, \mathbf{g}_o] &= \text{Split}_{3}(\mathbf{f}_{in}) \\
\mathbf{m}_t &= \sigma(\mathbf{g}_f) \odot \mathbf{m}_{t-1} + (1 - \sigma(\mathbf{g}_f)) \odot \tau(\mathbf{s}_t) \\
\mathbf{f}_{out} &= \sigma(\mathbf{g}_o) \odot \tau(\mathbf{m}_t) \cdot \mathbf{q}_{dec}
\end{aligned}
\end{equation}
where $\mathbf{F}_{t-1}$ is the temporal context, $\mathbf{q}_{dec}$ is a learnable decoder quantization step, $\mathbf{m}_t$ is the memory state, $\sigma(\cdot)$ and $\tau(\cdot)$ denote the gate and memory activation functions, and $\odot$ denotes element-wise multiplication.
The forget gate $\mathbf{g}_f$ controls information retention from the previous memory state, while the output gate $\mathbf{g}_o$ modulates the contribution to the output features.

Compared to the baseline decoder, the memory variant only triples the output channels of the final $1{\times}1$ convolution layer. The activation functions can be standard ($\sigma{=}\text{sigmoid}$, $\tau{=}\text{tanh}$) or piecewise-linear approximations using ReLU for hardware-friendly deployment.

Concurrent to our work, a recent study also introduced a memory mechanism for neural video compression \cite{msra_video_gan}.
In contrast, our design focuses on real-time, multi-platform deployment, emphasizing hardware compatibility and temporal robustness.

\paragraph{ReGLU.} 
\label{reglu}
The model performance is highly sensitive to activation function choice.
Moreover, optimal activation functions differ across architectures.
While cross-platform requirements restrict us to simple functions (ReLU, LeakyReLU), we can improve expressiveness through gating mechanisms that use only cross-platform-compatible operations.
Inspired by \cite{glu_variants}, we find that a channel split gating variant of ReGLU improves performance for the MLVC architecture without increasing encoder--decoder divergence:

\begin{equation}
\text{ReGLU}(\mathbf{x}) = \mathbf{x}_{:C/2} \odot \text{ReLU1}(\mathbf{x}_{C/2:})
\end{equation}
where $\mathbf{x}_{:C/2}$ and $\mathbf{x}_{C/2:}$ denote the first and second halves of channels, and $\text{ReLU1}(\cdot) = \min(\text{ReLU}(\cdot), 1)$ caps activations at 1. The capping stabilizes training by preventing large activations from the multiplicative gating.

We apply ReGLU in DCVC-RT's depth-wise convolution blocks, replacing the channel-wise addition with multiplicative gating, while using LeakyReLU elsewhere.

\paragraph{I-Frame Dropout.} To train a single model to handle both standard P-frames and I-frames without requiring a separate intra codec, we introduce I-frame dropout during training: with probability $p$, the model receives a constant gray frame in place of a true I-frame.
We use $p = 0.5$ in our experiments.
The effectiveness of this approach depends on the I-frame period: shorter periods benefit more, as I-frame conditions are encountered more frequently during inference.

\paragraph{Perceptual Losses.}
Cross-platform architectural constraints create a cumulative BD-rate penalty, and even with the above improvements, a gap remains compared to unconstrained architectures when measured by PSNR.
However, PSNR does not reflect subjective quality.
For real-world applications, perceptual optimization can matter more than pixel-wise accuracy.
Moreover, different applications have different perceptual priorities - video conferencing prioritizes facial detail, while screen sharing prioritizes text clarity.
We can narrow the perceptual quality gap by tailoring the codec to specific use cases through perceptual losses and region-of-interest (ROI) weighting.
Our subjective evaluation demonstrates that by adapting the codec to video conferencing, we achieve comparable results to reference DCVC-RT checkpoint despite lower PSNR.

Specifically, we use a perceptual loss with two components.
First, we use the common LPIPS \cite{lpips} loss to improve texture retention.
In addition, inspired by the success of \cite{variableroi, tlic} in the CLIC \cite{CLIC} challenge, we use a region of interest (ROI) mask to weight the loss pixels.
We use a face-segmentation model \cite{retinaface, farl} to derive the ROI mask during training.
We do not explicitly modulate latents with the mask like \cite{variableroi}, therefore avoiding dependence on an external model in deployment.
We also refrain from using a GAN loss as in \cite{agustsson2019generative, mentzer2020high, msra_video_gan}, since it can lead to temporal flickering and may introduce artifacts that perturb identity.
\section{Experiments}

\subsection{Experimental Setup}

\begin{table}[t]
\centering
\caption{Ablation study results showing PSNR-based BD-rate (\%) in YUV420 colorspace relative to HEVC-QSV anchor in same- and cross-platform scenarios.}
\label{tab:ablation}

\footnotesize
\setlength{\tabcolsep}{2.5pt}

\begin{tabular}{lccccccccccc}
\toprule
Arch & I & IP & RP & Act & SS & Mem & I-D & LTR & Avg (SP) & TH (SP) & TH (XP)\\
\midrule

DCVC-RT & \checkmark & -1 & 64 & WSiLU & \texttimes & \texttimes & \texttimes & \texttimes & -69.6 & - & $\infty$  \\
DCVC-RT & \texttimes & -1 & 64 & WSiLU & \texttimes & \texttimes & \texttimes & \texttimes & -68.6 & - & $\infty$\\
DCVC-RT & \texttimes & 64 & 0 & WSiLU & \texttimes & \texttimes & \texttimes & \texttimes & -59.4 & -55.9 & $\infty$\\

\midrule

0-prior & \texttimes & -1 & 64 & WSiLU & \texttimes & \texttimes & \texttimes & \texttimes & 20.5 & 9.6 & 718.1 \\

MLVC & \texttimes & -1 & 64 & WSiLU & \checkmark & \texttimes & \texttimes & \texttimes & -54.5 &  -56.1 & 101.8\\
MLVC & \texttimes & -1 & 64 & LReLU & \checkmark & \texttimes & \texttimes & \texttimes & -49.2 & -53.5 & 6.7\\
MLVC & \texttimes & -1 & 64 & LReLU & \checkmark & \checkmark & \texttimes & \texttimes & -52.1 & -58.4 & -21.9 \\
MLVC & \texttimes & -1 & 64 & ReGLU & \checkmark & \checkmark & \texttimes & \texttimes & -56.6 & -61.9 & -39.8 \\
MLVC & \texttimes & 64 & 0 & ReGLU & \checkmark & \checkmark & \texttimes & \texttimes & -41.1 & -42.7 & -41.0 \\
MLVC & \texttimes & 64 & 0 & ReGLU & \checkmark & \checkmark & \checkmark & \texttimes & -46.1 & -46.7 & -44.9 \\
MLVC & \texttimes & 1024 & 0 & ReGLU & \checkmark & \checkmark & \checkmark & \checkmark & -52.0 & -60.2 &  \textbf{-58.7} \\
MLVC-S & \texttimes & 1024 & 0 & ReGLU & \checkmark & \checkmark & \checkmark & \checkmark & -36.5 & -48.6 & -47.3\\

\bottomrule
\end{tabular}

\vspace{1mm}
\raggedright
\footnotesize{\textbf{Legend:}
   I: separate I-frame model,
   IP: intra period ($-1$: single intra frame),
   RP: reset period ($0$: no reset),
   SS: scale sending,
   Mem: memory module,
   I-D: I-frame dropout,
   LTR: long-term reference.
   SP: same-platform,
   XP: cross-platform,
   Avg: average BD-rate over HEVC B-E and 720p VCD sequences (NVIDIA GPU),
   TH: VCD TH subset at 360p (Apple M3),
   TH (SP): NPU encoder, NPU decoder,
   TH (XP): NPU encoder, GPU decoder,
   0-prior: no adaptive priors,
   MLVC-S: small variant of MLVC.}

\end{table}

\paragraph{Datasets.} We employ the same training schedule as DCVC-RT, with Vimeo-90K \cite{vimeo90k} septuplets used for first stage training, and longer Vimeo sequences up to 64 frames used for fine-tuning the model.
To assess the real-world deployability of neural codecs in video conferencing applications, we conduct a subjective test on the Video Conferencing Dataset (VCD) \cite{naderi2024vcd}. Objective metric results are also presented on the more common HEVC B-E \cite{hevc-data} datasets.

\paragraph{Training.}
We use the same variable rate control method as \cite{dcvc-rt} and train the model for the YUV colorspace. When training on long sequences, we use gradient and frame gradient clipping for improved stability. Additionally, we use gradient checkpointing in the fine-tuning stage. Training with sequences of up to 32 frames takes 5 days on 8 NVIDIA Tesla V100s. For perceptual models, we train the first stage using standard PSNR loss, but switch to the perceptual loss in fine-tuning.

\paragraph{Evaluation Metrics}

We measure the compression quality using BD-rate \cite{bjontegaard2001calculation}, computed from the actual encoded bitrate rather than the rate estimate from the entropy model. The distortion metric is PSNR in YUV420 colorspace or mean opinion score (MOS) in the subjective evaluation.
Additional metrics are provided in the supplementary material.
We primarily target 360p to 720p, as these resolutions are typical in real-time video conferencing.
For the HEVC datasets, metrics are reported at native resolution.

\paragraph{Cross-Platform Evaluation}
We report results in two settings: same-platform (SP), where encoder and decoder run on the same platform, and cross-platform (XP), where they run on different platforms (e.g., GPU encoder, NPU decoder). When cross-platform mismatch causes catastrophic decoding failure, we denote the BD-rate as $\infty$.

\paragraph{Anchor.}
We use Intel Quick Sync Video (HEVC-QSV)~\cite{QSV} as the single anchor for all BD-rate calculations throughout the paper, as it represents a widely deployed hardware codec comparable to what neural codecs must match in practice.
More details on the anchor configuration can be found in the supplementary material.
While we provide rate-distortion results for the reference implementation HM~\cite{HM} in the appendix, we omit VTM~\cite{VTM} and ECM~\cite{ECMref}, as they lack commodity hardware implementations.
We do compare against DCVC-RT, which outperforms HM, VTM, and ECM in PSNR.

\paragraph{P.910 Testing.}
We use the P.910 software package~\cite{naderi2024} to conduct subjective evaluations, running a 5-point ACR test as defined in ITU-T Recommendation P.910~\cite{P910}.
We collect 15 votes for each encoded clip in the VCD-TH subset.
All videos are displayed at 720p resolution to raters.
For lower processing resolutions (360p, 540p), the input is downscaled before encoding and upscaled back to 720p after decoding using ffmpeg's~\cite{ffmpeg} Lanczos algorithm.

\subsection{Experimental Results}

\paragraph{Ablation Study.}

Table~\ref{tab:ablation} presents an ablation study of architectural changes from DCVC-RT to MLVC. 
The original DCVC-RT achieves -69.6\% average BD-rate in the same-platform scenario, while running at 102 FPS at 360p resolution (Table~\ref{tab:bdrate}).
Our full model reaches -52\% BD-rate at 130 FPS at 360p, a gap of roughly 18 percentage points.
This gap is the direct cost of cross-platform constraints: scale sending, hardware-compatible activations, periodic I-frames, and a unified I/P-frame model each contribute (\cref{tab:ablation}).
Importantly, DCVC-RT produces catastrophic decoding failures on heterogeneous hardware and thus cannot be deployed across diverse platforms (BD-rate $= \infty$ in the XP column).
Among neural codecs that successfully decode across platforms, MLVC achieves the best BD-rate by a large margin ($-45.2$\% vs. $+15.0$\% for \cite{tian2024effortless} on HEVC B, see supplementary material for details).
LTR frames enable operation at much larger intra periods (IP=1024 vs.\ IP=64), providing additional BD-rate gains and improving robustness to packet loss. 
For higher resolutions, MLVC-S achieves 1080p encoding at 30\,FPS while still reaching $-37$\% BD-rate against the anchor.

\paragraph{Subjective Test.}

\begin{figure}[t]
  \centering

  \begin{subfigure}[t]{0.53\linewidth}
    \centering
    \includegraphics[width=\linewidth]{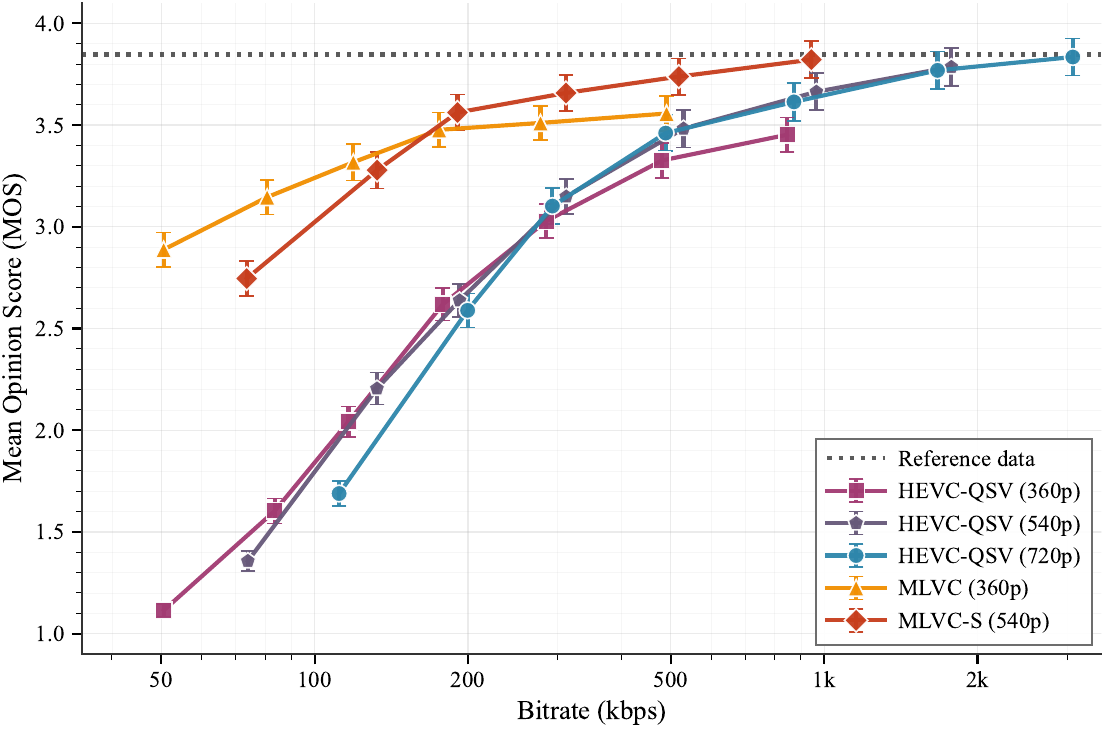}
    \caption{Rate–MOS curves.}
    \label{fig:rate_mos_vcd}
  \end{subfigure}
  \hfill
  \begin{subfigure}[t]{0.43\linewidth}
    \centering
    \includegraphics[width=\linewidth]{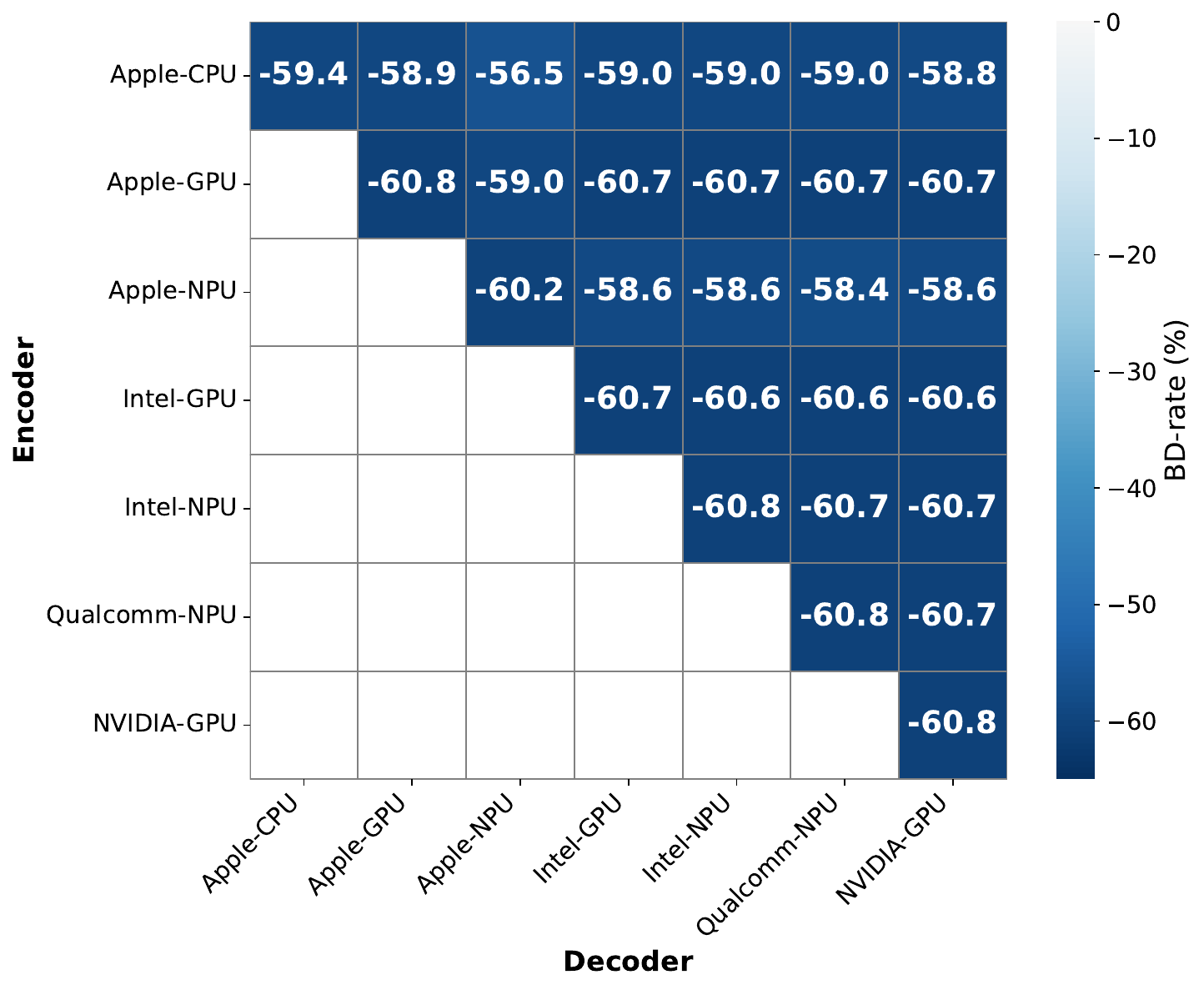}
    \caption{Cross-platform BD-rate matrix.}
    \label{fig:divergence_matrix}
  \end{subfigure}

  \caption{(a) Rate-MOS curves on VCD-TH, comparing hardware H.265 with the MLVC models at multiple processing resolutions. MLVC-S is a smaller model tuned for ~150 FPS at 540p. (b) Cross-platform BD-rate on VCD-TH 360p relative to HEVC-QSV. Rows indicate encoder platform, columns indicate decoder platform. Diagonal entries are same-platform; off-diagonal entries are cross-platform. Results demonstrate cross-platform compatibility.}
  \label{fig:performance_overview}
\end{figure}

We conduct a P.910~\cite{naderi2024} subjective evaluation on VCD-TH to assess MLVC in a video conferencing setting.
All codecs are tested at 360p and 540p processing resolutions, the H.265 hardware codec (HEVC-QSV) additionally at native 720p, and displayed at 720p. 
Results are shown in \cref{fig:rate_mos_vcd} and \cref{tab:bdrate}.

MLVC fine-tuned with perceptual losses achieves \textbf{--75.5\%} BD-rate (MOS) at 360p relative to the hardware HEVC-QSV anchor.
The lightweight MLVC-S achieves $-65.4$\% at 540p while matching HEVC-QSV in encoding speed.
Combining both via a bitrate ladder (MLVC at 360p for low bitrates, MLVC-S at 540p for high bitrates) yields $-71.8$\% overall.

The reference DCVC-RT is MSE-trained. To obtain a controlled perceptual baseline, we fine-tuned it with the same ROI and perceptual losses used for MLVC and evaluated the model at 360p.
Perceptual training improves DCVC-RT from $-76.4$\% to $-81.9$\% BD-rate (MOS), placing the cost of cross-platform constraints at roughly 6 percentage points relative to MLVC's $-75.5$\% --- a modest price given that DCVC-RT cannot decode across platforms.

\begin{table}[t]
\centering
\caption{BD-rate (MOS) relative to HEVC-QSV on VCD-TH at different processing resolutions: BD-360 and BD-540 denote BD-rate at 360p and 540p respectively, BD-All combines both via bitrate ladder. FPS indicates encoding speed on Apple M3 Pro NPU for neural codecs and Intel hardware for HEVC-QSV. MOS is always evaluated at 720p. CP denotes cross-platform capability.}
\label{tab:bdrate}

\footnotesize
\setlength{\tabcolsep}{4pt}
\renewcommand{\arraystretch}{1.1}

\begin{tabular}{lccccccc}
\toprule
Codec & IP & RP & BD-360 & BD-540 & BD-All & FPS (360/540) & CP \\
\midrule
HEVC-QSV   & -1 & –  &  0.0  &  0.0  &  0.0  & 221/170  & \checkmark \\
\midrule
DCVC-RT    & -1 & 64 & –76.4 & –80.3 & –77.9 & 102/46   & \texttimes \\
DCVC-RT (perceptual)    & -1 & 64 & –81.9 & – & – & 102/46   & \texttimes \\
\midrule
MLVC       & 64 & 32 & –75.5 & –78.8 & –77.4 & 130/66   & \checkmark \\
MLVC-S     & 64 & 32 & –     & –65.4 & –     & 300/152  & \checkmark \\
MLVC-multi & 64 & 32 & –75.5 & –65.4 & –71.8 & 130/152  & \checkmark \\
\bottomrule
\end{tabular}

\end{table}

\paragraph{Cross-Platform Validation.}
\Cref{fig:divergence_matrix} shows BD-rate for encoder-decoder platform pairs across Apple, Intel, Qualcomm, and NVIDIA hardware.
Most pairs achieve BD-rate within 2 points, confirming cross-platform compatibility. 
Importantly, all configurations avoid catastrophic failures.

\paragraph{Latency Evaluation.}
Table \ref{tab:speed_comparison} shows compute complexity results across resolutions on Apple M3 Pro ANE, Intel Lunar Lake NPU 4 i7/i9 version, and Qualcomm Snapdragon X Elite Hexagon NPU.
MLVC achieves real-time encoding and decoding (each ${>}30$\,FPS) up to 540p on all three platforms. MLVC-S extends real-time capability to 720p on all platforms and reaches 1080p at 30\,FPS on Apple M3 Pro.

\begin{table*}[t]
\centering
\scriptsize
\setlength{\tabcolsep}{3pt}
\renewcommand{\arraystretch}{1.05}
\caption{Speed comparison of MLVC models across different platforms and resolutions.}
\label{tab:speed_comparison}
\begin{tabular*}{\textwidth}{
@{\extracolsep{\fill}}
l l
r r  r r  r r  r r
r
r r
}
\toprule
\multirow{2}{*}{Model} & \multirow{2}{*}{Res.} &
\multicolumn{2}{c}{Apple (FPS)} & \multicolumn{2}{c}{Intel (FPS)} & \multicolumn{2}{c}{Qualcomm} & \multicolumn{2}{c}{Avg (FPS)} &
\multirow{2}{*}{\shortstack{Params \\ (M)}} &
\multicolumn{2}{c}{GFLOPs} \\
 & & Enc & Dec & Enc & Dec & Enc & Dec & Enc & Dec & & Enc & Dec \\
\midrule
MLVC      & 360p  & 129.5 & 122.9 & 98.0 & 98.0 & 80.0 & 77.5 & 103 & 99 & 18.3 & 63.9  & 66.2 \\
MLVC      & 540p  & 65.7  & 62.3  & 45.7 & 45.7 & 37.0 & 35.1 & 49 & 48 & 18.3 & 142.2 & 147.0 \\
MLVC      & 720p  & 33.8  & 33.1  & 24.4 & 26.5 & 21.4 & 20.4 & 27 & 27 & 18.3 & 253.8 & 262.5 \\
MLVC      & 1080p & 15.6  & 15.2  & 10.5 & 10.7 & 10.6 & 9.5 & 12 & 12 & 18.3 & 567.3 & 587.1 \\
\midrule
MLVC-S & 360p  & 300.3 & 277.8 & 263.2 & 270.3 & 193.4 & 173.9 & 252 & 241 & 5.4 & 21.6  & 23.2 \\
MLVC-S & 540p  & 152.0 & 137.6 & 120.5 & 125.0 & 89.6  & 77.9 & 121 & 114 & 5.4 & 48.0  & 51.4 \\
MLVC-S & 720p  & 83.8  & 73.5  & 62.1 & 70.4 & 51.3 & 44.1 & 66 & 63 & 5.4 & 85.7  & 91.9 \\
MLVC-S & 1080p & 39.5  & 34.6  & 24.0 & 25.8 & 26.5 & 19.6 & 30 & 27 & 5.4 & 191.9 & 205.7 \\
\bottomrule
\end{tabular*}
\end{table*}

\section{Conclusion}

We present MLVC, the first cross-platform neural video codec that achieves robust decoding across diverse consumer NPUs while maintaining competitive perceptual quality and real-time performance.
Our scale-sending mechanism eliminates catastrophic cross-platform decoding failures, while the memory module, ReGLU activations, I-frame dropout, and LTR frames reduce the compression cost of cross-platform constraints.
MLVC averages 100\,FPS at 360p across three major NPU platforms and achieves perceptual quality on par with original DCVC-RT on the video conferencing benchmark, while being 30\% faster.

Several directions remain for future work. Power consumption is likely to become the next optimization frontier now that real-time performance has been achieved. Additionally, simultaneous encoding and decoding on a single device at high resolutions (1080p and above) remains challenging for symmetric architectures and warrants further investigation.

%
%
\bibliographystyle{splncs04}
\bibliography{main}

\newif\ifsupp
\supptrue  

\ifsupp
\clearpage
\appendix
\clearpage
\appendix

\begin{center}
{\large \textbf{MLVC: Multi-platform Learned Video Codec for Real-World Deployment}}\\
{\large Supplementary Material}
\end{center}

\label{sec:supplementary}

This supplementary material provides additional experimental details (Section~\ref{sec:supp_experimental_setup}), extended ablations (Section~\ref{sec:supp_extended_results}), a formalization of our perceptual loss with ROI re-weighting (Section~\ref{sec:supp_perceptual_losses}), and a detailed analysis of cross-platform divergence (Section~\ref{sec:supp_divergence}).

\section{Experimental Setup}
\label{sec:supp_experimental_setup}

\subsection{Metrics and Colorspace}

All our metric calculations are performed in the YUV420 colorspace. Following \cite{dcvc-fm}, PSNR and MS-SSIM are computed for each channel separately and combined as $PSNR_{avg} = (6 \times PSNR_y + PSNR_u + PSNR_v) / 8$. Since LPIPS requires RGB inputs, we use BT.709 for conversion.

\subsection{Hardware Encoder Configuration}
We run the hardware encoder on a \href{https://en.wikipedia.org/wiki/Lunar_Lake}{Lunar Lake} processor (Intel(R) Core(TM) Ultra 9 288V) with the following ffmpeg \footnote{ffmpeg version 7.1.1-full\_build} config:

{\scriptsize
\begin{verbatim}
ffmpeg -init_hw_device qsv=hw
       -filter_hw_device hw
       -f rawvideo
       -pix_fmt yuv420p
       -s {width}x{height}
       -r {fps}
       -i {in_f}
       -c:v hevc_qsv
       -load_plugin hevc_hw
       -scenario 2
       -p_strategy 0
       -b_strategy 0
       -threads 1
       -sc_threshold 0
       -preset fast
       -bf 0
       -look_ahead 0
       -g 6000
       -i_qfactor 1.0
       -i_qoffset 0
       -b_qfactor 1.0
       -b_qoffset 0
       -refs 1
       -q {qp}
       {out_path}
\end{verbatim}
}

This configuration reflects typical settings used in modern real-time video conferencing applications, providing a practical balance between compression efficiency and encoding cost. 

\subsection{HEVC Anchors in Prior Work}

\begin{table}[t]
\centering
\scriptsize
\setlength{\tabcolsep}{2.5pt}

\caption{BD-rate comparison across HEVC anchors used in prior work on HEVC-B (all frames).
Assuming BD-rate transitivity, we convert reported results to the BD-rate relative
to our anchor for fair comparison across models (Rep.: reported BD-rate,
Impl.: implied BD-rate, IP: intra period).
}

\begin{tabular*}{\linewidth}{@{\extracolsep{\fill}}l r r l r r r c}
\toprule
\textbf{Anchor} &
\textbf{IP} &
\textbf{BD} &
\textbf{Used in} &
\makecell{\textbf{Model}\\\textbf{IP}} &
\makecell{\textbf{Rep.}\\\textbf{BD}} &
\makecell{\textbf{Impl.}\\\textbf{BD}} &
\makecell{\textbf{Cross}\\\textbf{plat.}} \\
\midrule

intel\_hw\_hevc (fast) & -1 & 0.0 & MLVC & 1024 & -45.2 & \textbf{-45.2} & \checkmark \\
ffmpeg (veryfast) & 10 & 39.6 & DVC~\cite{dvc} & 10 & 1.4 & 41.6 & \texttimes \\
ffmpeg (fast) & -1 & 7.1 & MobileCodec~\cite{mobilevnc} & 16 & 171.6 & 190.9 & \texttimes \\
ffmpeg (fast) & -1 & 7.1 & MobileNVC~\cite{mobilevnc} & 16 & 50.8 & 61.5 & \texttimes \\
ffmpeg (medium) & 12 & 21.3 & Effortless~\cite{tian2024effortless} & 32 & -5.2 & 15.0 & \checkmark \\
ffmpeg (veryslow) & 10 & 16.3 & DCVC~\cite{DCVC} & 10 & -26.0 & -13.9 & \texttimes \\
HM (encoder\_lowdelay\_main10) & -1 & -41.7 & DCVC-RT~\cite{dcvc-rt} & -1 & -43.5 & -67.0 & \texttimes \\

\midrule
\multicolumn{8}{c}{\textit{Additional anchors evaluated}} \\
\midrule

intel\_hw\_hevc (veryslow) & -1 & -7.5 & & & & & \\
ffmpeg (veryslow) & -1 & -3.3 & & & & & \\
ffmpeg (veryslow, psy-rd off) & -1 & -17.4 & & & & & \\
HM (encoder\_lowdelay\_P\_main) & -1 & -34.4 & & & & & \\

\bottomrule
\end{tabular*}

\label{tab:anchor_comparison}
\end{table}

As neural video codecs have evolved, the HEVC anchor configurations used in prior works have also changed.
For example, DVC used FFmpeg with preset \texttt{veryfast}, while DCVC changed it to \texttt{veryslow}, and DCVC-RT uses the official reference software configuration.
Table \ref{tab:anchor_comparison} summarizes the wide array of anchors used in relevant works cited in our paper.
The \textbf{BD} column reports BD-rate compared to our hardware HEVC anchor on HEVC-B.
The large variations reveal that anchor choices (preset, intra period, psycho-visual settings, and use of reference software) substantially affect reported BD-rates.
In particular, the intra period has a strong influence on compression efficiency.
While hardware encoders can be deployed with very long intra periods (e.g., IP=$-1$), many prior works use short intra periods, which make neural codecs appear more favorable in BD-rate comparisons.

Changing the Intel hardware encoder's preset from \texttt{fast} to \texttt{veryslow} has only a marginal impact on BD-rate, highlighting its fixed function nature.
With the \texttt{fast} preset, it is slightly better than x265.
However, prior work has used the default settings for x265, which enable psycho-visual optimization, diminishing PSNR.
When we disable this via \texttt{psy-rd=0:psy-rdoq=0:aq-mode=0:qcomp=1.0}, x265 performance increases substantially, as exemplified for the \texttt{veryslow} preset.
Finally, reference software HM achieves a significant improvement in BD-rate, while not being practical for deployment. 

To compare our model with prior work, we make use of the fact that BD-rate is almost always reported on the HEVC-B dataset.
By assuming the transitivity of BD-rates, we calculate \textbf{Implied BD} - the estimated BD-rate compared to our anchor.
While informative, this is a rough approximation, as differences in metric overlap intervals and interpolation methods (polynomial, piecewise or other) mean perfect transitivity does not hold.
Moreover, intra periods differ between models.
This fact itself often reflects architecture and training schedule improvements.

Previous approaches that have focused on mobile deployment \cite{mobilevnc} or cross-platform compatibility \cite{tian2024effortless} do not improve on our hardware HEVC anchor in BD-rate, making them infeasible for deployment.
In contrast, DCVC-RT beats our anchor by a wide margin.
MLVC also achieves large gains while additionally being cross-platform capable.

\subsection{Model Size}

We inherit the base configuration from DCVC-RT.
The recurrent feature dimensionality is reduced from 256 to 128, and the memory size is set to 128.
Based on early ablation studies, we use \(8\times\) spatial downsampling in the hyperprior and \(2\times\) channel sharing.
A more extensive ablation of scale sharing is provided in Section~\ref{sec:scale_sharing_ablation}.

For MLVC-S, the hyperencoder consists of a single layer that downsamples by \(4\times\).
The channel-sharing factor is increased to \(4\times\), the feature extractor is reduced to two layers, the number of channels is decreased from 256 to 192, and the \(y, z, \text{feature}\) channel dimensions are reduced from 128 to 48.

\section{Extended Results}
\label{sec:supp_extended_results}

\subsection{Additional Metrics}
Tables~\ref{tab:ablation-psnr}, \ref{tab:ablation-ms-ssim}, and \ref{tab:ablation-lpips} extend the ablation study from the main paper (Table~\ref{tab:ablation}) by reporting PSNR-, MS-SSIM-, and LPIPS-based BD-rates for each dataset individually. While Table~\ref{tab:ablation} reports the average PSNR-based BD-rate across datasets (Avg column), these tables expand that column into per-dataset results.
The conclusions remain consistent across metrics.

\newcommand{\ablationlegend}{%
\vspace{1mm}
\raggedright
\scriptsize{\textbf{Legend:}
I: separate I-frame model,
IP: intra period ($-1$=single intra frame),
RP: reset period ($0$=no reset),
Act: activation function,
SS: scale sending,
Mem: memory module,
I-D: I-frame dropout,
LTR: long-term reference,
TH/OB/BB/M: VCD subsets (Talking Head, Opaque Background, Background Blur, Mobile),
B-E: HEVC classes B-E,
0-prior: no adaptive priors,
MLVC-S: small MLVC variant.}}
  
\begin{table}[t]
\centering
\caption{Ablation study showing PSNR-based BD-rate (\%) in YUV420 colorspace on VCD and HEVC test sequences. Lower is better.}
\label{tab:ablation-psnr}
\scriptsize
\setlength{\tabcolsep}{2.5pt}
\resizebox{\textwidth}{!}{
\begin{tabular}{lccccccccccccccccc}
\toprule
Arch & I & IP & RP & Act & SS & Mem & I-D & LTR & TH & OB & BB & M & B & C & D & E & Avg \\
\midrule
DCVC-RT & \checkmark & -1 & 64 & WSiLU & \texttimes & \texttimes & \texttimes & \texttimes & -71.7 & -73.0 & -76.0 & -57.7 & -66.9 & -68.4 & -70.5 & -72.8 & -69.6 \\
DCVC-RT & \texttimes & -1 & 64 & WSiLU & \texttimes & \texttimes & \texttimes & \texttimes & -70.3 & -71.4 & -75.1 & -57.0 & -66.0 & -67.8 & -69.9 & -71.5 & -68.6 \\
DCVC-RT & \texttimes & 64 & 0 & WSiLU & \texttimes & \texttimes & \texttimes & \texttimes & -59.1 & -59.3 & -68.4 & -51.7 & -59.1 & -62.6 & -64.2 & -50.9 & -59.4 \\
0-prior & \texttimes & -1 & 64 & WSiLU & \texttimes & \texttimes & \texttimes & \texttimes & 20.3 & 39.1 & 35.2 & 21.0 & -10.8 & -29.3 & -39.6 & 127.8 & 20.5 \\
MLVC & \texttimes & -1 & 64 & WSiLU & \checkmark & \texttimes & \texttimes & \texttimes & -57.5 & -55.8 & -62.5 & -46.4 & -51.9 & -53.2 & -56.5 & -52.4 & -54.5 \\
MLVC & \texttimes & -1 & 64 & LReLU & \checkmark & \texttimes & \texttimes & \texttimes & -54.4 & -52.8 & -59.4 & -39.4 & -43.3 & -44.9 & -48.4 & -50.7 & -49.2 \\
MLVC & \texttimes & -1 & 64 & LReLU & \checkmark & \checkmark & \texttimes & \texttimes & -58.1 & -59.3 & -60.6 & -39.8 & -44.9 & -46.7 & -47.7 & -59.9 & -52.1 \\
MLVC & \texttimes & -1 & 64 & ReGLU & \checkmark & \checkmark & \texttimes & \texttimes & -60.5 & -60.6 & -62.9 & -44.0 & -52.3 & -54.1 & -56.5 & -62.2 & -56.6 \\
MLVC & \texttimes & 64 & 0 & ReGLU & \checkmark & \checkmark & \texttimes & \texttimes & -41.3 & -38.6 & -53.0 & -36.3 & -41.0 & -46.0 & -47.0 & -25.8 & -41.1 \\
MLVC & \texttimes & 64 & 0 & ReGLU & \checkmark & \checkmark & \checkmark & \texttimes & -46.7 & -45.7 & -57.5 & -39.7 & -44.7 & -48.3 & -49.7 & -36.4 & -46.1 \\
MLVC & \texttimes & 1024 & 0 & ReGLU & \checkmark & \checkmark & \checkmark & \checkmark & -57.9 & -59.4 & -58.4 & -39.5 & -45.2 & -48.3 & -48.2 & -59.3 & -52.0 \\
MLVC-S & \texttimes & 1024 & 0 & ReGLU & \checkmark & \checkmark & \checkmark & \checkmark & -43.1 & -45.7 & -44.2 & -17.7 & -24.0 & -33.0 & -34.7 & -49.7 & -36.5 \\
\bottomrule
\end{tabular}
}

\ablationlegend

\end{table}

\begin{table}[t]
\centering
\caption{Ablation study showing MS-SSIM-based BD-rate (\%) in YUV420 colorspace on VCD and HEVC test sequences. Lower is better.}
\label{tab:ablation-ms-ssim}
\scriptsize
\setlength{\tabcolsep}{2.5pt}
\resizebox{\textwidth}{!}{
\begin{tabular}{lccccccccccccccccc}
\toprule
Arch & I & IP & RP & Act & SS & Mem & I-D & LTR & TH & OB & BB & M & B & C & D & E & Avg \\
\midrule
DCVC-RT & \checkmark & -1 & 64 & WSiLU & \texttimes & \texttimes & \texttimes & \texttimes & -73.4 & -73.7 & -80.9 & -59.1 & -69.8 & -73.6 & -75.8 & -73.0 & -72.4 \\
DCVC-RT & \texttimes & -1 & 64 & WSiLU & \texttimes & \texttimes & \texttimes & \texttimes & -71.1 & -71.2 & -79.9 & -57.8 & -68.0 & -72.9 & -75.1 & -70.9 & -70.9 \\
DCVC-RT & \texttimes & 64 & 0 & WSiLU & \texttimes & \texttimes & \texttimes & \texttimes & -56.9 & -55.0 & -72.7 & -50.8 & -59.5 & -66.9 & -69.0 & -41.8 & -59.1 \\
0-prior & \texttimes & -1 & 64 & WSiLU & \texttimes & \texttimes & \texttimes & \texttimes & 9.6 & 27.8 & 1.3 & 20.4 & -17.3 & -41.8 & -51.4 & 104.1 & 6.6 \\
MLVC & \texttimes & -1 & 64 & WSiLU & \checkmark & \texttimes & \texttimes & \texttimes & -60.1 & -56.4 & -69.9 & -48.3 & -57.4 & -62.9 & -66.7 & -56.3 & -59.7 \\
MLVC & \texttimes & -1 & 64 & LReLU & \checkmark & \texttimes & \texttimes & \texttimes & -57.8 & -55.1 & -67.2 & -43.9 & -51.8 & -57.7 & -61.8 & -55.4 & -56.3 \\
MLVC & \texttimes & -1 & 64 & LReLU & \checkmark & \checkmark & \texttimes & \texttimes & -60.6 & -60.1 & -68.2 & -44.7 & -53.5 & -59.0 & -62.8 & -60.8 & -58.7 \\
MLVC & \texttimes & -1 & 64 & ReGLU & \checkmark & \checkmark & \texttimes & \texttimes & -61.9 & -60.1 & -69.3 & -46.5 & -57.2 & -62.5 & -66.0 & -61.3 & -60.6 \\
MLVC & \texttimes & 64 & 0 & ReGLU & \checkmark & \checkmark & \texttimes & \texttimes & -41.1 & -34.7 & -59.6 & -38.0 & -46.0 & -54.8 & -57.8 & -19.3 & -43.9 \\
MLVC & \texttimes & 64 & 0 & ReGLU & \checkmark & \checkmark & \checkmark & \texttimes & -48.8 & -44.9 & -65.8 & -42.7 & -51.2 & -57.3 & -60.9 & -33.9 & -50.7 \\
MLVC & \texttimes & 1024 & 0 & ReGLU & \checkmark & \checkmark & \checkmark & \checkmark & -59.3 & -58.4 & -65.9 & -41.3 & -50.2 & -56.2 & -58.4 & -55.1 & -55.6 \\
MLVC-S & \texttimes & 1024 & 0 & ReGLU & \checkmark & \checkmark & \checkmark & \checkmark & -45.8 & -45.6 & -54.3 & -19.3 & -33.3 & -45.7 & -51.2 & -46.6 & -42.7 \\
\bottomrule
\end{tabular}
}

\ablationlegend
\end{table}

\begin{table}[t]
\centering
\caption{Ablation study showing LPIPS-based BD-rate (\%) in YUV420 colorspace on VCD and HEVC test sequences. Lower is better.}
\label{tab:ablation-lpips}
\scriptsize
\setlength{\tabcolsep}{2.5pt}
\resizebox{\textwidth}{!}{
\begin{tabular}{lccccccccccccccccc}
\toprule
Arch & I & IP & RP & Act & SS & Mem & I-D & LTR & TH & OB & BB & M & B & C & D & E & Avg \\
\midrule
DCVC-RT & \checkmark & -1 & 64 & WSiLU & \texttimes & \texttimes & \texttimes & \texttimes & -65.5 & -59.1 & -64.0 & -37.1 & -43.8 & -62.4 & -69.3 & -53.7 & -56.9 \\
DCVC-RT & \texttimes & -1 & 64 & WSiLU & \texttimes & \texttimes & \texttimes & \texttimes & -63.2 & -55.5 & -63.7 & -34.6 & -40.9 & -61.9 & -68.2 & -49.7 & -54.7 \\
DCVC-RT & \texttimes & 64 & 0 & WSiLU & \texttimes & \texttimes & \texttimes & \texttimes & -45.6 & -30.8 & -54.3 & -22.9 & -25.3 & -53.9 & -59.9 & -3.4 & -37.0 \\
0-prior & \texttimes & -1 & 64 & WSiLU & \texttimes & \texttimes & \texttimes & \texttimes & 30.9 & 61.1 & 80.6 & 55.8 & 34.0 & -20.7 & -37.5 & 199.4 & 50.5 \\
MLVC & \texttimes & -1 & 64 & WSiLU & \checkmark & \texttimes & \texttimes & \texttimes & -51.8 & -35.7 & -55.1 & -24.1 & -25.9 & -47.6 & -57.1 & -33.4 & -41.3 \\
MLVC & \texttimes & -1 & 64 & LReLU & \checkmark & \texttimes & \texttimes & \texttimes & -49.6 & -34.9 & -53.3 & -16.7 & -11.6 & -38.6 & -49.5 & -34.1 & -36.0 \\
MLVC & \texttimes & -1 & 64 & LReLU & \checkmark & \checkmark & \texttimes & \texttimes & -54.5 & -45.9 & -53.8 & -18.8 & -15.5 & -41.0 & -50.8 & -40.1 & -40.0 \\
MLVC & \texttimes & -1 & 64 & ReGLU & \checkmark & \checkmark & \texttimes & \texttimes & -55.9 & -43.7 & -54.7 & -23.8 & -23.6 & -48.7 & -57.0 & -38.2 & -43.2 \\
MLVC & \texttimes & 64 & 0 & ReGLU & \checkmark & \checkmark & \texttimes & \texttimes & -30.6 & -6.4 & -40.6 & -11.8 & -4.4 & -38.5 & -47.2 & 25.5 & -19.3 \\
MLVC & \texttimes & 64 & 0 & ReGLU & \checkmark & \checkmark & \checkmark & \texttimes & -40.9 & -20.3 & -45.5 & -21.1 & -16.3 & -44.2 & -52.9 & 1.9 & -29.9 \\
MLVC & \texttimes & 1024 & 0 & ReGLU & \checkmark & \checkmark & \checkmark & \checkmark & -54.2 & -41.2 & -50.2 & -17.3 & -15.4 & -42.4 & -49.0 & -35.1 & -38.1 \\
MLVC-S & \texttimes & 1024 & 0 & ReGLU & \checkmark & \checkmark & \checkmark & \checkmark & -37.8 & -20.7 & -36.9 & 12.8 & 13.9 & -26.1 & -36.7 & -20.8 & -19.0 \\
\bottomrule
\end{tabular}
}

\ablationlegend
\end{table}

\subsection{Additional Benchmarks: UVG and MCL-JCV}
\label{sec:supp_uvg_mcljcv}

The main paper uses VCD and HEVC test sequences as the primary benchmarks. To further evaluate generalization beyond these main benchmarks, we add two additional high-resolution video compression benchmarks: UVG and MCL-JCV. Both datasets contain 1080p sequences and are evaluated at their native resolution. Table~\ref{tab:supp_generalization} reports PSNR BD-rate in YUV420 relative to the same HEVC-QSV anchor used in the main paper.

\begin{table}[t]
\centering
\small
\setlength{\tabcolsep}{4.5pt}
\caption{Additional PSNR BD-rate (\%) results relative to HEVC-QSV.}
\label{tab:supp_generalization}
\begin{tabular}{@{}lrr@{}}
\toprule
Dataset & DCVC-RT & MLVC\\
\midrule
Original average & $-69.6$ & $-52.0$\\
UVG (1080p) & $-69.4$ & $-51.7$ \\
MCL-JCV (1080p) & $-63.0$ & $-47.3$ \\
Average incl. UVG/MCL-JCV & $-68.9$ & $-51.5$ \\
\midrule
VCD-TH (cross-platform scenario) & $\infty$ (fail) & $-58.7$ \\
\bottomrule
\end{tabular}
\end{table}

MLVC maintains large gains over hardware HEVC on both additional 1080p datasets, achieving $-51.7\%$ BD-rate on UVG and $-47.3\%$ on MCL-JCV.
The gap to DCVC-RT remains consistent across datasets.
This gap reflects the cost of eliminating the same-platform bit-exactness assumption: DCVC-RT provides a same-platform upper bound, while MLVC is designed for reliable deployment across heterogeneous encoder-decoder platforms.

\subsection{Latency Breakdown}
\label{sec:supp_latency_breakdown}

We provide a latency breakdown to isolate the per-frame speed effect of each architectural change from DCVC-RT to MLVC. Following the architectural columns in Table~\ref{tab:ablation}, Table~\ref{tab:supp_latency_breakdown} reports encoder and decoder FPS on the Apple M3 Pro NPU at 360p. I-frame dropout and LTR are training/inference-time policies that do not affect per-frame compute, so they are omitted from this breakdown.

\begin{table}[t]
\centering
\small
\setlength{\tabcolsep}{4pt}
\caption{Latency breakdown on Apple M3 Pro NPU at 360p.}
\label{tab:supp_latency_breakdown}
\begin{tabular}{lcccrr}
\toprule
Arch & Act & SS & Mem & Enc FPS & Dec FPS\\
\midrule
DCVC-RT & WSiLU & \texttimes & \texttimes & 102 & 75 \\
MLVC & WSiLU & \checkmark & \texttimes & 102 & 97 \\
MLVC & LReLU & \checkmark & \texttimes & 130 & 123 \\
MLVC & LReLU & \checkmark & \checkmark & 131 & 123 \\
MLVC & ReGLU & \checkmark & \checkmark & 130 & 123 \\
MLVC-S & ReGLU & \checkmark & \checkmark & 300 & 278 \\
\bottomrule
\end{tabular}
\end{table}

Scale sending improves decoder speed, increasing decoder FPS from 75 to 97 without changing encoder FPS. Hardware-compatible activations improve both encoder and decoder speed, increasing FPS from 102/97 to 130/123. ReGLU matches the FPS of LReLU, indicating that the gating adds no measurable runtime cost in this setting. The memory row has essentially unchanged FPS by design: we keep the recurrent state size constant by replacing 256 reference-feature channels with 128 reference-feature channels plus 128 memory channels.

\subsection{Scale Sharing Ablation}
\label{sec:scale_sharing_ablation}
We ablate the spatial and channel sharing factors in the scale sharing mechanism.
Spatial sharing is increased by adding additional 2x downsampling blocks in the hyperprior.
Figure~\ref{fig:scale_ablation} reports the resulting BD-rate, averaged over VCD 720p and HEVC sequences.
The experiments in the main paper used three 2x downsampling blocks in the hyperprior, and a channel sharing factor of two.
Further improvements may be achieved by increasing the channel sharing factor.

\begin{figure}[t]
  \centering
   \includegraphics[width=0.8\linewidth]{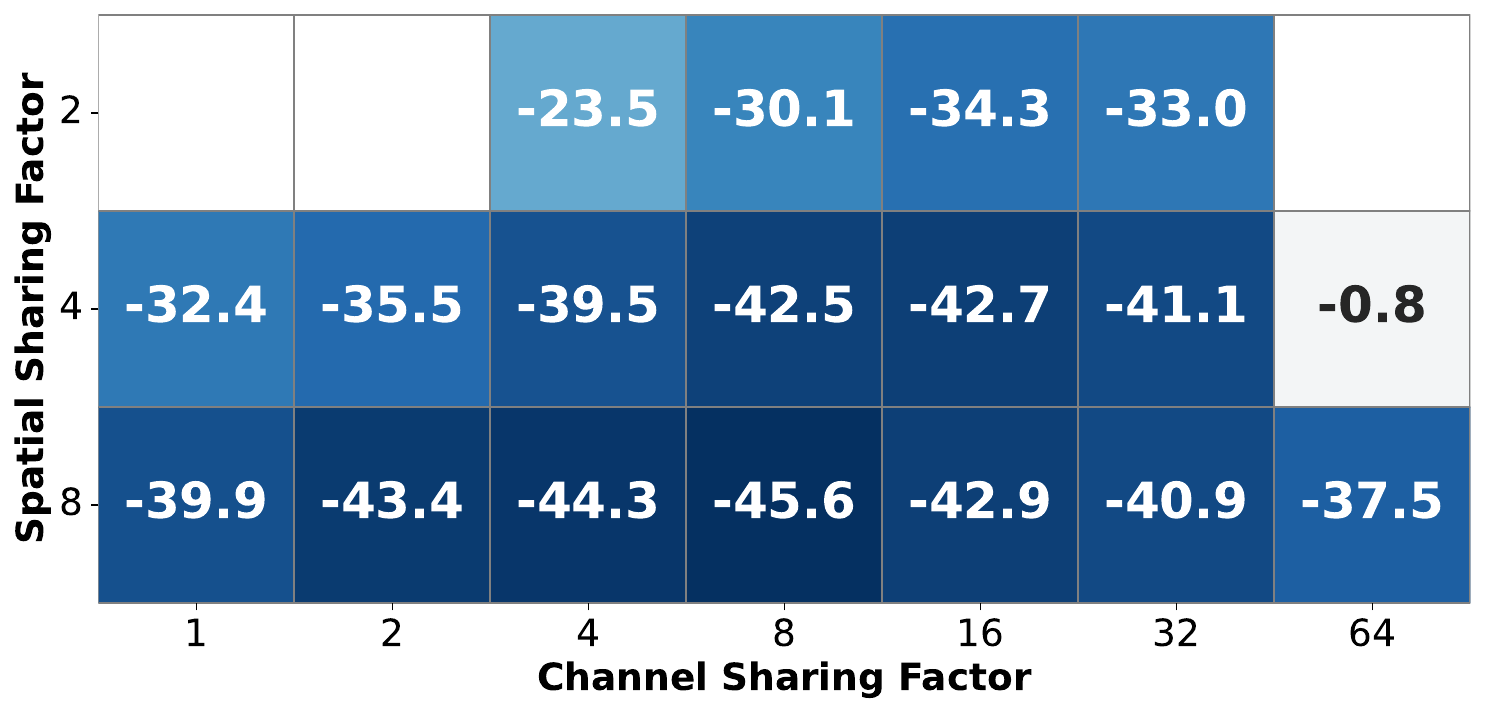}

   \caption{Ablation study on spatial and channel sharing factors in the scale sharing mechanism. BD-rate is averaged over VCD 720p and HEVC sequences. All models are trained on sequences of up to 32 frames.}
   \label{fig:scale_ablation}
\end{figure}

\subsection{Activation Functions}

We evaluate the impact of different activation functions.
Specifically, we compare LeakyReLU, WSiLU, SiLU, GELU, HardSwish and ReGLU activations while keeping all other components fixed.
For ReGLU, as explained in \cref{reglu}, we only change the activation in the depth-wise convolution block by replacing channel addition with multiplicative gating and using ReLU1, while using LeakyReLU elsewhere.
For all other activations, we change all activation functions throughout the network.
The results are shown in Table \ref{tab:act_ablation}.
As discussed in Section~\ref{sec:act_func_divergence}, some activations give better BD-rate but are unsuitable for cross-platform deployment due to large numeric divergence between devices.

\begin{table*}[t]
\centering
\caption{Activation function ablation study results showing BD-rate (PSNR) in YUV420 colorspace on VCD and HEVC test sequences.}
\label{tab:act_ablation}
\begin{tabular}{lccccccccc}
\toprule
Activation & TH & TH-OB & TH-BB & TH-M & HEVC B & HEVC C & HEVC D & HEVC E & Avg \\
\midrule
WSiLU & -61.7 & -60.9 & -63.7 & -46.5 & -52.8 & -54.6 & -56.7 & -63.1 & -57.5 \\
ReGLU & -60.5 & -60.6 & -62.9 & -44.0 & -52.3 & -54.1 & -56.5 & -62.2 & -56.6 \\
GELU & -59.8 & -60.0 & -62.6 & -44.5 & -51.4 & -53.1 & -55.0 & -61.8 & -56.0 \\
SiLU & -59.1 & -60.6 & -61.9 & -41.6 & -47.4 & -49.7 & -51.2 & -63.3 & -54.3 \\
LReLU & -58.1 & -59.3 & -60.6 & -39.8 & -44.9 & -46.7 & -47.7 & -59.9 & -52.1 \\
HardSwish & -57.1 & -59.0 & -60.3 & -40.4 & -45.5 & -45.4 & -47.2 & -61.4 & -52.0 \\
\bottomrule
\end{tabular}

\vspace{1mm}
\raggedright
\end{table*}

\subsection{Scaling Laws}
While our main results are presented for MLVC/MLVC-S running at 100-120 FPS, we observe across our experiments significant scaling laws between frame coding time in ms and BD-rate, as shown in Figure \ref{fig:scaling_law}. This trend indicates that as more powerful NPU hardware becomes available, larger model variants can be deployed within the same latency budget, enabling improved compression performance even without future architectural innovations.

\begin{figure}[t]
  \centering
   \includegraphics[width=0.9\linewidth]{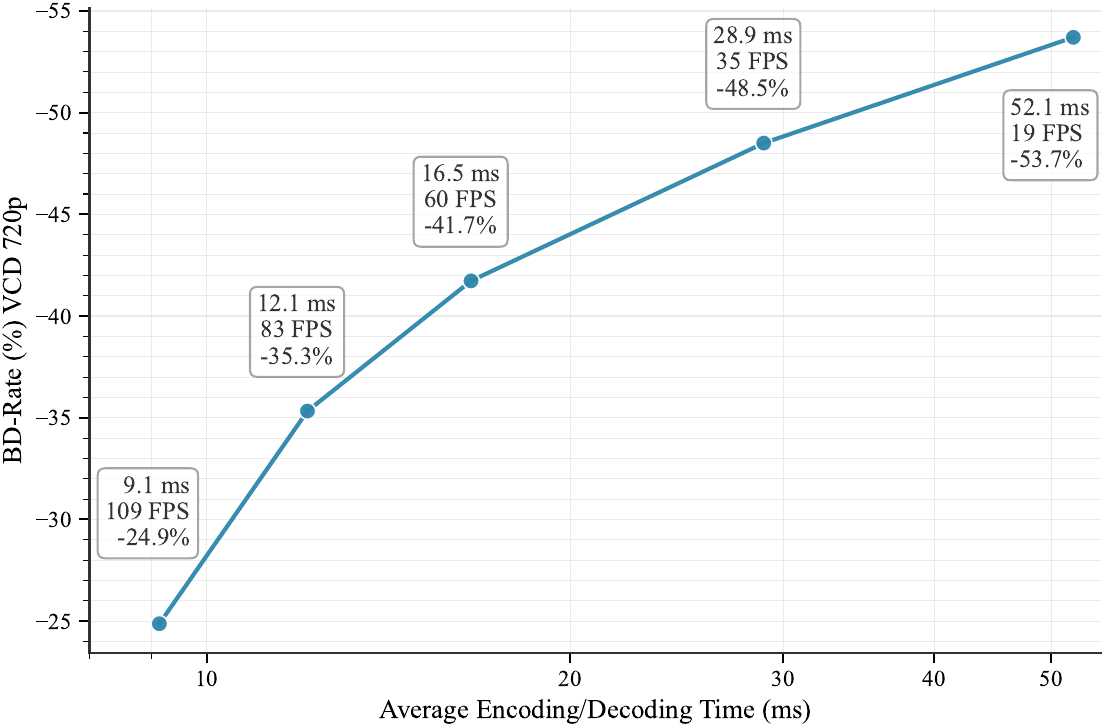}

   \caption{Rate-distortion performance improves with additional computational resources. Each point represents a different model configuration with varying computational complexity. Annotations show the average encoding/decoding latency (ms), throughput (FPS) and BD-rate (PSNR) relative to the HEVC-QSV on VCD 720p. MLVC benefits directly from hardware advances: as NPUs improve each year, MLVC provides effortless performance gains for neural video compression without algorithmic changes.}
   \label{fig:scaling_law}
\end{figure}

\subsection{Cross-platform LTR Ablation}

To evaluate the effectiveness of the Long-Term Reference (LTR) mechanism in a true cross-platform setting, we measure BD-rate degradation when encoding and decoding are performed on different compute units.
Table~\ref{tab:bdrate_comparison} reports the BD-rate increase (Delta) when moving from the same compute unit (Apple M3 GPU$\rightarrow$GPU) to a heterogeneous configuration (Apple M3 GPU$\rightarrow$NPU).

As illustrated in Fig.~\ref{fig:divergence-example}, this GPU–NPU pair exhibits one of the largest numerical divergences among the tested device combinations.
At comparable levels of cross-platform divergence (Delta), the LTR mechanism allows the use of longer intra periods (IP), which improves the BD-rate.

\begin{table}[t]
\setlength{\tabcolsep}{5pt}
\centering
\small
\caption{Cross-platform LTR comparison between same-device (Apple M3 GPU-GPU) and cross-device decoding (Apple M3 GPU-NPU). Delta shows the BD-rate degradation from cross-device decoding. At comparable Delta, LTR enables longer intra periods and thus better BD-rate.}
\label{tab:bdrate_comparison}
\begin{tabular}{cccccc}
\toprule
\textbf{IP} & \textbf{LTR} & \textbf{LTR-Period} & \multicolumn{2}{c}{\textbf{BD-rate}} & \textbf{Delta} \\
 & & & \scriptsize GPU-GPU & \scriptsize GPU-NPU & \\
\midrule
32 & No & 0 & -10.2\% & -9.0\% & -1.2 \\
64 & No & 0 & -39.9\% & -37.8\% & -2.2 \\
128 & No & 0 & -51.7\% & -47.0\% & -4.7 \\
\midrule
64 & Yes & 32 & -34.8\% & -33.8\% & -1.0 \\
128 & Yes & 32 & -42.4\% & -41.4\% & -1.0 \\
128 & Yes & 64 & -48.4\% & -46.4\% & -2.0 \\
256 & Yes & 32 & -45.1\% & -44.1\% & -1.0 \\
256 & Yes & 64 & -51.7\% & -49.8\% & -1.9 \\
256 & Yes & 128 & -55.5\% & -50.8\% & -4.7 \\
\bottomrule
\end{tabular}
\end{table}

\subsection{LTR under Scene Changes}
\label{sec:supp_scene_changes}

We evaluate MLVC under scene changes by constructing a scene-change set from MCL-JCV. Specifically, we concatenate 15 video pairs with fixed cuts, using 64 frames before and 64 frames after each cut. This setup tests the behavior of long-term reference (LTR) prediction when the stored reference becomes stale after an abrupt cut.

MLVC uses a unified I/P-frame model: an I-frame is encoded as a P-frame whose reference is a gray frame. Therefore, the model is explicitly trained to handle large reference/current-frame mismatch. At IP=$64$, explicit I-frame refresh achieves $-52.3\%$ BD-rate, while LTR with period 64 achieves $-50.3\%$. Thus, using a stale LTR reference incurs a $2.0$ percentage-point BD-rate penalty under abrupt scene changes. This penalty can be avoided in deployment by standard scene-cut detection that triggers an I-frame refresh. LTR remains preferable for long continuous prediction chains, as shown in the IP=$1024$ setting in Table~\ref{tab:ablation}.

\subsection{Rate-Distortion Curves}
To complement the BD-rate metrics reported in Table~\ref{tab:ablation}, we provide full rate-distortion curves in Figure~\ref{fig:ablation_rd_curves} for three key model configurations from our ablation study: DCVC-RT (our non-cross-platform starting point), MLVC (our proposed model) and MLVC-S (the lightweight version).

\subsection{Full-frame Visual Comparisons}
Figure~\ref{fig:frame_comparison_full} provides the full frames from which the crops in the main paper's Figure \ref{fig:frame_comparison_zoomed} were extracted.
The full frames show that the quality improvements extend across entire images, not just in the ROI (face) regions.

\begin{figure*}[t]
  \centering
  \includegraphics[width=\linewidth]{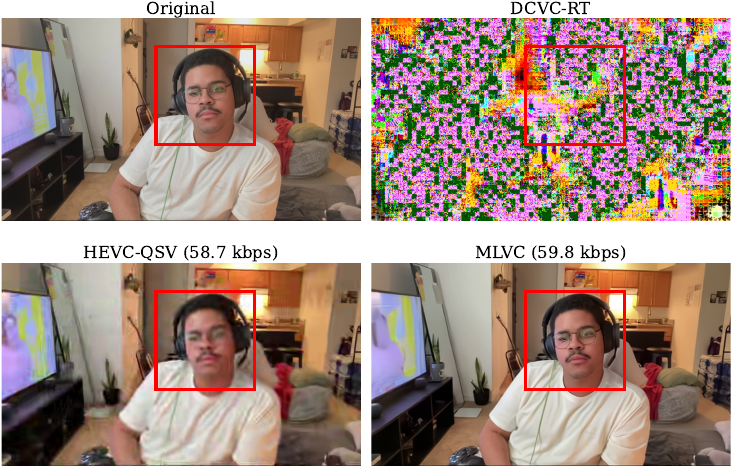}
  \caption{Full frames used for the close-up comparison in Figure \ref{fig:frame_comparison_zoomed}.}
  \label{fig:frame_comparison_full}
\end{figure*}

\subsection{Decoder Divergence in Long Prediction Chains}

As discussed in \cref{sec:reducing-divergence}, small platform-dependent numerical differences can accumulate over long prediction chains and lead to gradual divergence between encoder and decoder states. \Cref{fig:divergence-example} shows an extreme example of this phenomenon.

\begin{figure}[t]
  \centering
   \includegraphics[width=1.0\linewidth]{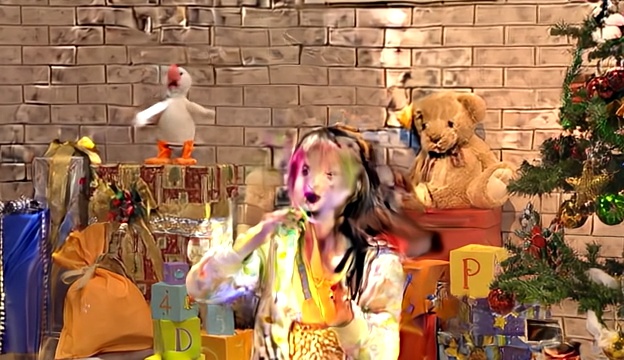}
   \caption{Decoder divergence in a long prediction chain. While the scale-sending mechanism eliminates catastrophic failures, platform-dependent numerical differences lead to accumulating prediction errors. The image illustrates an extreme case, where features from the teddy bear gradually bleed into the face region after a long chain of P-frame predictions from a single I-frame when using a model with WSiLU activation.}
   \label{fig:divergence-example}
\end{figure}

\section{Perceptual losses}
\label{sec:supp_perceptual_losses}

To make use of the benefits of both the perceptual loss \cite{lpips} and region of interest weighting, we use the following distortion loss formulation:

\begin{align*}
L = w_m \cdot \overline{M \otimes \text{MSE}(x, \hat{x})} + w_l \cdot \overline{M \otimes \text{LPIPS}(x, \hat{x})} \\
M_i = \begin{cases}
w_{ROI} & \text{if } i \in \text{ROI} \\
w_{BG} & \text{otherwise}
\end{cases} \\
w_{ROI} = \frac{k}{p(1+k)}, \quad w_{BG} = \frac{1}{(1-p)(1+k)}
\end{align*}
where MSE and LPIPS denote their single channel spatial versions, $\{k, w_m, w_l\}$ are hyperparameters, $p$ signifies the training dataset specific probability of a pixel belonging to the ROI, and $M$ is the 2D weight mask with elements according to $M_i$.

The weights $w_{ROI}$ and $w_{BG}$ are derived to satisfy two constraints: a) we want to keep roughly the same magnitude of total loss, to not affect the rate-distortion trade-off, b) the hyperparameter $k$ should correspond to the relative importance of ROI pixels to background pixels. 

Formally average spatial error $\overline{M \otimes E(x, \hat{x})}$ for an image is defined as 

\begin{align*}
\ell = \frac{1}{N} \sum_{i=1}^{N} e_i \cdot w_i = \frac{1}{N} \sum_{i \in ROI} e_i \cdot w_{ROI} + \frac{1}{N} \sum_{i \in BG} e_i \cdot w_{BG}
\end{align*}

However, $N$ can be redefined to mean all pixels in a dataset, rather than a single image. Then, given the expectation

\begin{align*}
\mathbb{E}[\ell] &= \frac{1}{N} \cdot p \cdot N \cdot w_{ROI} \cdot \mathbb{E}[e_i] \\
&\quad + \frac{1}{N} \cdot (1-p) \cdot N \cdot w_{BG} \cdot \mathbb{E}[e_i]
\end{align*}
we can enforce ROI pixels to contribute $k$ times more than the background by having

\begin{align}
\frac{\mathbb{E}[\ell_{ROI}]}{\mathbb{E}[\ell_{BG}]} = \frac{p \cdot w_{ROI}}{(1-p) \cdot w_{BG}} \stackrel{!}{=} k
\label{eq:ratio}
\end{align}
and also enforcing the constraint from a):
\begin{align}
(1-p) \cdot w_{BG} + p \cdot w_{ROI} = 1
\label{eq:magnitude}
\end{align}
Solving \eqref{eq:ratio} and \eqref{eq:magnitude} for $w_{ROI}$ and $w_{BG}$ yields the results.

Parametrizing $p$ allows us to keep a similar ROI and background trade-off (given by $k$) even when using different training datasets or different segmentation masks.

\section{Cross-Platform Divergence Analysis}
\label{sec:supp_divergence}

\subsection{Quantized Networks}

\subsubsection{INT8 Quantization}

\definecolor{c0}{RGB}{130,210,130}
\definecolor{c1}{RGB}{200,238,178}
\definecolor{c2}{RGB}{220,240,170}
\definecolor{c3}{RGB}{255,230,150}
\definecolor{c4}{RGB}{255,210,140}
\definecolor{c5}{RGB}{255,190,150}
\definecolor{c6}{RGB}{255,170,170}

\begin{table}[t]
\centering
\scriptsize
\setlength{\tabcolsep}{2.5pt}
\caption{Standard deviation of output error for INT8-quantized convolution across devices and runtimes.}
\label{tab:quant_conv}

\resizebox{\linewidth}{!}{%
\begin{tabular}{llllccccccccccccccc}
\toprule
 &  &  &  & \multicolumn{9}{c}{CPU} & \multicolumn{2}{c}{GPU} & \multicolumn{4}{c}{NPU} \\
\cmidrule(lr){5-13}\cmidrule(lr){14-15}\cmidrule(lr){16-19}
 &  &  &  & \multicolumn{5}{c}{FP32} & \multicolumn{4}{c}{FP16} & FP16 & FP32 & \multicolumn{2}{c}{FP16} & \multicolumn{2}{c}{FP32} \\
\cmidrule(lr){5-9}\cmidrule(lr){10-13}\cmidrule(lr){16-17}\cmidrule(lr){18-19}
 &  &  &  & \multicolumn{3}{c}{ORT} & CoreML & OpenVINO & \multicolumn{2}{c}{ORT} & CoreML & OpenVINO & CoreML &  & CoreML & OpenVINO &  & ORT-QNN \\
\cmidrule(lr){5-7}\cmidrule(lr){10-11}
Device & Precision & Runtime & Vendor & Intel & Apple & Qualcomm & Apple & Intel & Apple & Intel & Apple & Intel & Apple &  & Apple & Intel &  & Qualcomm \\
\midrule
\multirow{9}{*}{CPU} & \multirow{5}{*}{FP32} & \multirow{3}{*}{ORT} & Intel & \cellcolor{c0}0.00000 & \cellcolor{c0}0.00000 & \cellcolor{c0}0.00000 & \cellcolor{c1}0.00002 & \cellcolor{c1}0.00001 & \cellcolor{c6}0.01980 & \cellcolor{c6}0.01980 & \cellcolor{c5}0.00354 & \cellcolor{c1}0.00001 & \cellcolor{c4}0.00177 & \cellcolor{c2}0.00011 & \cellcolor{c5}0.00919 & \cellcolor{c3}0.00036 & \cellcolor{c3}0.00036 & \cellcolor{c5}0.00914 \\
 &  &  & Apple & \cellcolor{c0}0.00000 & \cellcolor{c0}0.00000 & \cellcolor{c0}0.00000 & \cellcolor{c1}0.00002 & \cellcolor{c1}0.00001 & \cellcolor{c6}0.01980 & \cellcolor{c6}0.01980 & \cellcolor{c5}0.00354 & \cellcolor{c1}0.00001 & \cellcolor{c4}0.00177 & \cellcolor{c2}0.00011 & \cellcolor{c5}0.00919 & \cellcolor{c3}0.00036 & \cellcolor{c3}0.00036 & \cellcolor{c5}0.00914 \\
 &  &  & Qualcomm & \cellcolor{c0}0.00000 & \cellcolor{c0}0.00000 & \cellcolor{c0}0.00000 & \cellcolor{c1}0.00002 & \cellcolor{c1}0.00001 & \cellcolor{c6}0.01980 & \cellcolor{c6}0.01980 & \cellcolor{c5}0.00354 & \cellcolor{c1}0.00001 & \cellcolor{c4}0.00177 & \cellcolor{c2}0.00011 & \cellcolor{c5}0.00919 & \cellcolor{c3}0.00036 & \cellcolor{c3}0.00036 & \cellcolor{c5}0.00914 \\
 &  & CoreML & Apple & \cellcolor{c1}0.00002 & \cellcolor{c1}0.00002 & \cellcolor{c1}0.00002 & \cellcolor{c0}0.00000 & \cellcolor{c1}0.00002 & \cellcolor{c6}0.01980 & \cellcolor{c6}0.01980 & \cellcolor{c5}0.00354 & \cellcolor{c1}0.00002 & \cellcolor{c4}0.00177 & \cellcolor{c2}0.00011 & \cellcolor{c5}0.00919 & \cellcolor{c3}0.00036 & \cellcolor{c3}0.00036 & \cellcolor{c5}0.00914 \\
 &  & OpenVINO & Intel & \cellcolor{c1}0.00001 & \cellcolor{c1}0.00001 & \cellcolor{c1}0.00001 & \cellcolor{c1}0.00002 & \cellcolor{c0}0.00000 & \cellcolor{c6}0.01980 & \cellcolor{c6}0.01980 & \cellcolor{c5}0.00354 & \cellcolor{c0}0.00000 & \cellcolor{c4}0.00177 & \cellcolor{c2}0.00011 & \cellcolor{c5}0.00919 & \cellcolor{c3}0.00036 & \cellcolor{c3}0.00036 & \cellcolor{c5}0.00914 \\
\cmidrule(lr){2-4}
 & \multirow{4}{*}{FP16} & \multirow{2}{*}{ORT} & Apple & \cellcolor{c6}0.01980 & \cellcolor{c6}0.01980 & \cellcolor{c6}0.01980 & \cellcolor{c6}0.01980 & \cellcolor{c6}0.01980 & \cellcolor{c0}0.00000 & \cellcolor{c0}0.00000 & \cellcolor{c6}0.01991 & \cellcolor{c6}0.01980 & \cellcolor{c6}0.01990 & \cellcolor{c6}0.01980 & \cellcolor{c6}0.02060 & \cellcolor{c6}0.01979 & \cellcolor{c6}0.01979 & \cellcolor{c6}0.02140 \\
 &  &  & Intel & \cellcolor{c6}0.01980 & \cellcolor{c6}0.01980 & \cellcolor{c6}0.01980 & \cellcolor{c6}0.01980 & \cellcolor{c6}0.01980 & \cellcolor{c0}0.00000 & \cellcolor{c0}0.00000 & \cellcolor{c6}0.01991 & \cellcolor{c6}0.01980 & \cellcolor{c6}0.01990 & \cellcolor{c6}0.01980 & \cellcolor{c6}0.02060 & \cellcolor{c6}0.01979 & \cellcolor{c6}0.01979 & \cellcolor{c6}0.02140 \\
 &  & CoreML & Apple & \cellcolor{c5}0.00354 & \cellcolor{c5}0.00354 & \cellcolor{c5}0.00354 & \cellcolor{c5}0.00354 & \cellcolor{c5}0.00354 & \cellcolor{c6}0.01991 & \cellcolor{c6}0.01991 & \cellcolor{c0}0.00000 & \cellcolor{c5}0.00354 & \cellcolor{c5}0.00348 & \cellcolor{c5}0.00354 & \cellcolor{c5}0.00924 & \cellcolor{c5}0.00354 & \cellcolor{c5}0.00354 & \cellcolor{c5}0.00936 \\
 &  & OpenVINO & Intel & \cellcolor{c1}0.00001 & \cellcolor{c1}0.00001 & \cellcolor{c1}0.00001 & \cellcolor{c1}0.00002 & \cellcolor{c0}0.00000 & \cellcolor{c6}0.01980 & \cellcolor{c6}0.01980 & \cellcolor{c5}0.00354 & \cellcolor{c0}0.00000 & \cellcolor{c4}0.00177 & \cellcolor{c2}0.00011 & \cellcolor{c5}0.00919 & \cellcolor{c3}0.00036 & \cellcolor{c3}0.00036 & \cellcolor{c5}0.00914 \\
\midrule
\multirow{2}{*}{GPU} & \multirow{1}{*}{FP16} & CoreML & Apple & \cellcolor{c4}0.00177 & \cellcolor{c4}0.00177 & \cellcolor{c4}0.00177 & \cellcolor{c4}0.00177 & \cellcolor{c4}0.00177 & \cellcolor{c6}0.01990 & \cellcolor{c6}0.01990 & \cellcolor{c5}0.00348 & \cellcolor{c4}0.00177 & \cellcolor{c0}0.00000 & \cellcolor{c4}0.00177 & \cellcolor{c5}0.00917 & \cellcolor{c4}0.00179 & \cellcolor{c4}0.00179 & \cellcolor{c5}0.00915 \\
\cmidrule(lr){2-4}
 & \multirow{1}{*}{FP32} & CoreML & Apple & \cellcolor{c2}0.00011 & \cellcolor{c2}0.00011 & \cellcolor{c2}0.00011 & \cellcolor{c2}0.00011 & \cellcolor{c2}0.00011 & \cellcolor{c6}0.01980 & \cellcolor{c6}0.01980 & \cellcolor{c5}0.00354 & \cellcolor{c2}0.00011 & \cellcolor{c4}0.00177 & \cellcolor{c0}0.00000 & \cellcolor{c5}0.00919 & \cellcolor{c3}0.00037 & \cellcolor{c3}0.00037 & \cellcolor{c5}0.00914 \\
\midrule
\multirow{4}{*}{NPU} & \multirow{2}{*}{FP16} & CoreML & Apple & \cellcolor{c5}0.00919 & \cellcolor{c5}0.00919 & \cellcolor{c5}0.00919 & \cellcolor{c5}0.00919 & \cellcolor{c5}0.00919 & \cellcolor{c6}0.02060 & \cellcolor{c6}0.02060 & \cellcolor{c5}0.00924 & \cellcolor{c5}0.00919 & \cellcolor{c5}0.00917 & \cellcolor{c5}0.00919 & \cellcolor{c0}0.00000 & \cellcolor{c5}0.00919 & \cellcolor{c5}0.00919 & \cellcolor{c6}0.01198 \\
 &  & OpenVINO & Intel & \cellcolor{c3}0.00036 & \cellcolor{c3}0.00036 & \cellcolor{c3}0.00036 & \cellcolor{c3}0.00036 & \cellcolor{c3}0.00036 & \cellcolor{c6}0.01979 & \cellcolor{c6}0.01979 & \cellcolor{c5}0.00354 & \cellcolor{c3}0.00036 & \cellcolor{c4}0.00179 & \cellcolor{c3}0.00037 & \cellcolor{c5}0.00919 & \cellcolor{c0}0.00000 & \cellcolor{c0}0.00000 & \cellcolor{c5}0.00914 \\
\cmidrule(lr){2-4}
 & \multirow{2}{*}{FP32} & OpenVINO & Intel & \cellcolor{c3}0.00036 & \cellcolor{c3}0.00036 & \cellcolor{c3}0.00036 & \cellcolor{c3}0.00036 & \cellcolor{c3}0.00036 & \cellcolor{c6}0.01979 & \cellcolor{c6}0.01979 & \cellcolor{c5}0.00354 & \cellcolor{c3}0.00036 & \cellcolor{c4}0.00179 & \cellcolor{c3}0.00037 & \cellcolor{c5}0.00919 & \cellcolor{c0}0.00000 & \cellcolor{c0}0.00000 & \cellcolor{c5}0.00914 \\
 &  & ORT-QNN & Qualcomm & \cellcolor{c5}0.00914 & \cellcolor{c5}0.00914 & \cellcolor{c5}0.00914 & \cellcolor{c5}0.00914 & \cellcolor{c5}0.00914 & \cellcolor{c6}0.02140 & \cellcolor{c6}0.02140 & \cellcolor{c5}0.00936 & \cellcolor{c5}0.00914 & \cellcolor{c5}0.00915 & \cellcolor{c5}0.00914 & \cellcolor{c6}0.01198 & \cellcolor{c5}0.00914 & \cellcolor{c5}0.00914 & \cellcolor{c0}0.00000 \\
\bottomrule
\end{tabular}}
\end{table}

\definecolor{c0}{RGB}{130,210,130}
\definecolor{c1}{RGB}{215,240,175}
\definecolor{c2}{RGB}{255,230,150}
\definecolor{c3}{RGB}{255,170,170}

\begin{table}[t]
\centering
\scriptsize
\setlength{\tabcolsep}{2.5pt}
\caption{Standard deviation of output error for FP16 convolution across devices and runtimes.}
\label{tab:fp16_conv}

\resizebox{\linewidth}{!}{%
\begin{tabular}{lllcccccccccc}
\toprule
\multicolumn{3}{c}{FP16 conv} & \multicolumn{5}{c}{CPU} & \multicolumn{2}{c}{GPU} & \multicolumn{3}{c}{NPU} \\
\cmidrule(lr){4-8}\cmidrule(lr){9-10}\cmidrule(lr){11-13}
 &  &  & \multicolumn{3}{c}{ORT} & CoreML & OpenVINO & CoreML & OpenVINO & CoreML & OpenVINO & ORT-QNN \\
\cmidrule(lr){4-6}
Device & Runtime & Vendor & Apple & Intel & Qualcomm & Apple & Intel & Apple & Intel & Apple & Intel & Qualcomm \\
\midrule
\multirow{5}{*}{CPU} & \multirow{3}{*}{ORT} & Apple & \cellcolor{c0}0.00000 & \cellcolor{c0}0.00000 & \cellcolor{c3}0.00098 & \cellcolor{c3}0.00098 & \cellcolor{c0}0.00000 & \cellcolor{c0}0.00000 & \cellcolor{c1}0.00001 & \cellcolor{c2}0.00012 & \cellcolor{c1}0.00001 & \cellcolor{c1}0.00001 \\
 &  & Intel & \cellcolor{c0}0.00000 & \cellcolor{c0}0.00000 & \cellcolor{c3}0.00098 & \cellcolor{c3}0.00098 & \cellcolor{c0}0.00000 & \cellcolor{c0}0.00000 & \cellcolor{c1}0.00001 & \cellcolor{c2}0.00012 & \cellcolor{c1}0.00001 & \cellcolor{c1}0.00001 \\
 &  & Qualcomm & \cellcolor{c3}0.00098 & \cellcolor{c3}0.00098 & \cellcolor{c0}0.00000 & \cellcolor{c0}0.00000 & \cellcolor{c3}0.00098 & \cellcolor{c3}0.00098 & \cellcolor{c3}0.00098 & \cellcolor{c3}0.00098 & \cellcolor{c3}0.00098 & \cellcolor{c3}0.00098 \\
 & \multirow{1}{*}{CoreML} & Apple & \cellcolor{c3}0.00098 & \cellcolor{c3}0.00098 & \cellcolor{c0}0.00000 & \cellcolor{c0}0.00000 & \cellcolor{c3}0.00098 & \cellcolor{c3}0.00098 & \cellcolor{c3}0.00098 & \cellcolor{c3}0.00098 & \cellcolor{c3}0.00098 & \cellcolor{c3}0.00098 \\
 & \multirow{1}{*}{OpenVINO} & Intel & \cellcolor{c0}0.00000 & \cellcolor{c0}0.00000 & \cellcolor{c3}0.00098 & \cellcolor{c3}0.00098 & \cellcolor{c0}0.00000 & \cellcolor{c0}0.00000 & \cellcolor{c1}0.00001 & \cellcolor{c2}0.00012 & \cellcolor{c1}0.00001 & \cellcolor{c1}0.00001 \\
\midrule
\multirow{2}{*}{GPU} & \multirow{1}{*}{CoreML} & Apple & \cellcolor{c0}0.00000 & \cellcolor{c0}0.00000 & \cellcolor{c3}0.00098 & \cellcolor{c3}0.00098 & \cellcolor{c0}0.00000 & \cellcolor{c0}0.00000 & \cellcolor{c1}0.00001 & \cellcolor{c2}0.00012 & \cellcolor{c1}0.00001 & \cellcolor{c1}0.00001 \\
 & \multirow{1}{*}{OpenVINO} & Intel & \cellcolor{c1}0.00001 & \cellcolor{c1}0.00001 & \cellcolor{c3}0.00098 & \cellcolor{c3}0.00098 & \cellcolor{c1}0.00001 & \cellcolor{c1}0.00001 & \cellcolor{c0}0.00000 & \cellcolor{c2}0.00012 & \cellcolor{c1}0.00001 & \cellcolor{c1}0.00000 \\
\midrule
\multirow{3}{*}{NPU} & \multirow{1}{*}{CoreML} & Apple & \cellcolor{c2}0.00012 & \cellcolor{c2}0.00012 & \cellcolor{c3}0.00098 & \cellcolor{c3}0.00098 & \cellcolor{c2}0.00012 & \cellcolor{c2}0.00012 & \cellcolor{c2}0.00012 & \cellcolor{c0}0.00000 & \cellcolor{c2}0.00012 & \cellcolor{c2}0.00012 \\
 & \multirow{1}{*}{OpenVINO} & Intel & \cellcolor{c1}0.00001 & \cellcolor{c1}0.00001 & \cellcolor{c3}0.00098 & \cellcolor{c3}0.00098 & \cellcolor{c1}0.00001 & \cellcolor{c1}0.00001 & \cellcolor{c1}0.00001 & \cellcolor{c2}0.00012 & \cellcolor{c0}0.00000 & \cellcolor{c1}0.00001 \\
 & \multirow{1}{*}{ORT-QNN} & Qualcomm & \cellcolor{c1}0.00001 & \cellcolor{c1}0.00001 & \cellcolor{c3}0.00098 & \cellcolor{c3}0.00098 & \cellcolor{c1}0.00001 & \cellcolor{c1}0.00001 & \cellcolor{c1}0.00000 & \cellcolor{c2}0.00012 & \cellcolor{c1}0.00001 & \cellcolor{c0}0.00000 \\
\bottomrule
\end{tabular}}
\end{table}

INT8 quantization of the model is expected to eliminate cross-device divergence, since integer calculations are generally guaranteed to produce identical results across platforms. To verify this hypothesis, we constructed a simple PyTorch test model consisting of a pointwise convolution with 128 channels, random weights, and zero bias (the simplest case). The model was quantized to INT8 using the standard static quantization scheme provided by PyTorch with the QNNPACK backend. The resulting INT8-quantized model was exported to Core ML, OpenVINO, and ONNX Runtime (ORT) formats to enable testing on Apple, Intel, and Qualcomm devices. All models were exported in two precisions: FP16 and FP32. In quantized models, this only affects the data type of the quantization scales and the input/output tensors. Special care was taken to ensure that all models used identical static quantization parameters, weights, and input values, regardless of precision. For all three formats, the exported model shared a similar structure:  
\texttt{Quantize → DeQuantize → Conv → Quantize → DeQuantize}.
This is the standard QDQ (Quantize--DeQuantize) representation used for quantized models. When executed on-device, the model is compiled to use a quantized convolution, resulting in the structure:  
\texttt{Quantize → QuantizedConv → DeQuantize}.  

The model was tested on Apple, Intel, and Qualcomm devices using Core ML, OpenVINO, and ORT with both CPU and Qualcomm AI Engine Direct (QNN) execution providers. Random inputs of shape $(128 \times 256 \times 256)$ drawn from a standard normal distribution were passed through the model across different devices, runtimes, and precisions. The results, shown in Table \ref{tab:quant_conv}, report the standard deviation of the output error for all combinations of device, runtime, precision, and vendor. Even for such a simple model the results show divergence across most combinations. For the FP32 model on CPU, consistency is highest: outputs from ORT across different platforms align closely; however, both Core ML and OpenVINO already exhibit slight deviations for FP32 on CPU. For the FP16 models on CPU, only Intel OpenVINO matches the FP32 CPU results; all other configurations show significant errors, with ORT being the most divergent. The results for GPU and NPU configurations deviate significantly from those on CPU, and overall, no pair of GPUs or NPUs produces identical outputs. FP16 testing on the Qualcomm device was not possible due to ORT-QNN limitations; however, NPUs typically support only FP16 operations. This is evident from Intel NPU results, which are identical for FP16 and FP32.

\subsubsection{Fake Quantization}

\definecolor{good}{RGB}{180,235,180}
\definecolor{mid}{RGB}{255,230,150}
\definecolor{bad}{RGB}{255,170,170}
  
\begin{table}[t]
\centering
\footnotesize
\setlength{\tabcolsep}{4pt}
\caption{Output error from fake quantization under different rounding modes and precisions.}
\label{tab:fake_quant}

\begin{tabular}{llllcccc}
\toprule
\multicolumn{4}{c}{Fake Quantization} & \multicolumn{2}{c}{Round} & \multicolumn{2}{c}{Truncate} \\
\cmidrule(lr){5-6}\cmidrule(lr){7-8}
Device & Precision & Runtime & Vendor & FP16 & FP32 & FP16 & FP32 \\
\midrule

\multirow{9}{*}{CPU}
& \multirow{5}{*}{FP32}
& ORT      & Apple    & \cellcolor{mid}0.00298 & \cellcolor{good}0.00000 & \cellcolor{bad}0.02164 & \cellcolor{bad}0.02180 \\
&          & ORT      & Intel    & \cellcolor{mid}0.00298 & \cellcolor{good}0.00000 & \cellcolor{bad}0.02164 & \cellcolor{bad}0.02180 \\
&          & ORT      & Qualcomm & \cellcolor{mid}0.00298 & \cellcolor{good}0.00000 & \cellcolor{bad}0.02164 & \cellcolor{bad}0.02180 \\
&          & CoreML   & Apple    & \cellcolor{mid}0.00298 & \cellcolor{good}0.00000 & \cellcolor{bad}0.02164 & \cellcolor{bad}0.02180 \\
&          & OpenVINO & Intel    & \cellcolor{mid}0.00298 & \cellcolor{good}0.00000 & \cellcolor{bad}0.02164 & \cellcolor{bad}0.02180 \\
\cmidrule(lr){2-8}
& \multirow{4}{*}{FP16}
& ORT      & Apple    & \cellcolor{bad}0.02179 & \cellcolor{bad}0.02180 & \cellcolor{mid}0.00262 & \cellcolor{good}0.00000 \\
&          & ORT      & Intel    & \cellcolor{bad}0.02179 & \cellcolor{bad}0.02180 & \cellcolor{mid}0.00262 & \cellcolor{good}0.00000 \\
&          & CoreML   & Apple    & \cellcolor{mid}0.00298 & \cellcolor{good}0.00000 & \cellcolor{bad}0.02164 & \cellcolor{bad}0.02180 \\
&          & OpenVINO & Intel    & \cellcolor{mid}0.00298 & \cellcolor{good}0.00000 & \cellcolor{bad}0.02164 & \cellcolor{bad}0.02180 \\
\midrule

\multirow{2}{*}{GPU}
& FP16 & CoreML & Apple & \cellcolor{mid}0.00299 & \cellcolor{mid}0.00013 & \cellcolor{bad}0.02164 & \cellcolor{bad}0.02180 \\
& FP32 & CoreML & Apple & \cellcolor{mid}0.00299 & \cellcolor{mid}0.00013 & \cellcolor{bad}0.02164 & \cellcolor{bad}0.02180 \\
\midrule

\multirow{4}{*}{NPU}
& \multirow{2}{*}{FP16}
& CoreML   & Apple    & \cellcolor{mid}0.00617 & \cellcolor{mid}0.00612 & \cellcolor{bad}0.02189 & \cellcolor{bad}0.02205 \\
&          & OpenVINO & Intel    & \cellcolor{mid}0.00298 & \cellcolor{good}0.00000 & \cellcolor{bad}0.02164 & \cellcolor{bad}0.02180 \\
\cmidrule(lr){2-8}
& \multirow{2}{*}{FP32}
& OpenVINO & Intel    & \cellcolor{mid}0.00298 & \cellcolor{good}0.00000 & \cellcolor{bad}0.02164 & \cellcolor{bad}0.02180 \\
&          & ORT-QNN  & Qualcomm & \cellcolor{mid}0.00894 & \cellcolor{mid}0.00894 & \cellcolor{bad}0.01759 & \cellcolor{bad}0.01778 \\
\bottomrule
\end{tabular}
\end{table}

To better understand the source of divergence, we conducted additional tests. The divergence might originate from the quantization of the input or the dequantization of the output. To isolate this effect, we performed another set of experiments using a model that contained only a single fake quantization layer. This fake quantization operation can be expressed as:

\begin{equation}
\text{Quantized} = \operatorname{clip}\!\Big( \operatorname{int}\big( \operatorname{round}\big( \frac{\text{input}}{\text{scale}} \big) \big)\Big)
\end{equation}

\begin{equation}
\text{Dequantized} = \text{scale} \times \text{Quantized}
\end{equation}

However, the rounding step can be implemented in different ways, such as round, truncate, or even custom bit-shifting operations. In addition, these calculations may occur at different precisions or by using different rounding rules, which further contributes to variability. We tested two rounding functions (round and truncate) under FP16 and FP32 precisions. The results are shown in Table \ref{tab:fake_quant}. It is evident that on CPU, most inference engines use the standard round function along with FP32 precision. One notable exception is ORT, which appears to use truncate instead of round for FP16 models; this likely explains the large divergence observed in the INT8 quantized convolution results in Table \ref{tab:quant_conv}. On GPUs, the behavior is close to FP32 round but not an exact match. On NPUs, Intel uses FP32 with round for both FP16 and FP32 models, whereas Apple NPU does not match any ground truth in the table, although its behavior is closer to round than to truncate. On Qualcomm, the results are incorrect because ORT-QNN seems to optimize away this simple fake quantization layer.

\subsubsection{IEEE 754 Adherence}

\newcommand{\cmark}{\ding{51}}
\newcommand{\xmark}{\ding{55}}

\definecolor{good}{RGB}{200,235,200}     
\definecolor{partial}{RGB}{255,235,150}  
\definecolor{approx}{RGB}{255,200,150}   
\definecolor{bad}{RGB}{255,160,160}      

\begin{table}[t]
\centering
\footnotesize
\setlength{\tabcolsep}{4pt}

\caption{Compliance of FP16 arithmetic and rounding operations with IEEE~754 across devices and runtimes.}
\label{tab:fp16_ops}

\begin{tabular}{lllccccccc}
\toprule
Device & Runtime & Vendor &
Add & Sub & Mul & Recip & Round & Floor & Clip \\
\midrule

\multirow{5}{*}{CPU}
 & ORT
 & Apple    & \cellcolor{good}\cmark & \cellcolor{good}\cmark & \cellcolor{good}\cmark & \cellcolor{good}\cmark & \cellcolor{good}\cmark & \cellcolor{good}\cmark & \cellcolor{good}\cmark \\
 & ORT
 & Intel    & \cellcolor{good}\cmark & \cellcolor{good}\cmark & \cellcolor{good}\cmark & \cellcolor{good}\cmark & \cellcolor{good}\cmark & \cellcolor{good}\cmark & \cellcolor{good}\cmark \\
 & ORT
 & Qualcomm & \cellcolor{good}\cmark & \cellcolor{good}\cmark & \cellcolor{good}\cmark & \cellcolor{good}\cmark & \cellcolor{good}\cmark & \cellcolor{good}\cmark & \cellcolor{good}\cmark \\

 & CoreML
 & Apple    & \cellcolor{good}\cmark & \cellcolor{good}\cmark & \cellcolor{good}\cmark & \cellcolor{bad}\xmark  & \cellcolor{good}\cmark & \cellcolor{good}\cmark & \cellcolor{good}\cmark \\

 & OpenVINO
 & Intel    & \cellcolor{good}\cmark & \cellcolor{good}\cmark & \cellcolor{good}\cmark & \cellcolor{good}\cmark & \cellcolor{good}\cmark & \cellcolor{good}\cmark & \cellcolor{good}\cmark \\

\midrule

\multirow{2}{*}{GPU}
 & CoreML
 & Apple    & \cellcolor{good}\cmark & \cellcolor{good}\cmark & \cellcolor{good}\cmark & \cellcolor{bad}\xmark  & \cellcolor{bad}\xmark  & \cellcolor{good}\cmark & \cellcolor{good}\cmark \\

 & OpenVINO
 & Intel    & \cellcolor{good}\cmark & \cellcolor{good}\cmark & \cellcolor{good}\cmark & \cellcolor{good}\cmark & \cellcolor{good}\cmark & \cellcolor{good}\cmark & \cellcolor{good}\cmark \\

\midrule

\multirow{3}{*}{NPU}
 & CoreML
 & Apple    & \cellcolor{partial}\cmark & \cellcolor{partial}\cmark & \cellcolor{approx}\xmark & \cellcolor{bad}\xmark  & \cellcolor{bad}\xmark  & \cellcolor{good}\cmark & \cellcolor{good}\cmark \\

 & OpenVINO
 & Intel    & \cellcolor{good}\cmark & \cellcolor{good}\cmark & \cellcolor{approx}\xmark & \cellcolor{good}\cmark & \cellcolor{good}\cmark & \cellcolor{good}\cmark & \cellcolor{good}\cmark \\

 & ORT-QNN
 & Qualcomm & \cellcolor{good}\cmark & \cellcolor{partial}\cmark & \cellcolor{partial}\cmark & \cellcolor{bad}\xmark  & \cellcolor{bad}\xmark  & \cellcolor{good}\cmark & \cellcolor{good}\cmark \\

\bottomrule
\end{tabular}
\end{table}

To further investigate the sources of divergence, we conducted additional tests on basic algebraic operations and compared the results against IEEE 754 floating-point calculations. In this test, we limited ourselves to FP16 precision. Even within the IEEE 754 standard, discrepancies can occur because the standard defines five rounding modes: \textit{to nearest, ties to even} (default for binary floats), \textit{to nearest, ties away from zero}, \textit{toward $0$}, \textit{toward $+\infty$}, and \textit{toward $-\infty$}. The results are shown in Table \ref{tab:fp16_ops}.
Most runtime and device combinations fully adhere to IEEE 754 with the default \textit{to nearest, ties to even} rounding mode on CPU and GPU (except Core ML for reciprocal). However, the situation for NPUs is quite different. Apple appears to use \textit{to nearest, ties away from zero} for addition and subtraction. For multiplication and reciprocal, the results do not strictly follow IEEE 754, although multiplication is close to \textit{to nearest, ties away from zero}. Intel NPU generally follows the standard with the default rounding mode, except for multiplication, which is close but not exact. Qualcomm NPU seems to use different rounding rules for addition and subtraction and a non-default mode for multiplication; reciprocal results are also non-standard.
In addition to basic arithmetic operations, we compared the results of round, floor, and clip against expected values. While clip and floor are consistently implemented across all devices, runtimes, and vendors, the round operation is inconsistent on GPUs and NPUs. It should be noted that some runtimes allow configuration of rounding modes, but for NPUs these options are limited, especially in Core ML. Given that all NPUs in Table \ref{tab:fp16_ops} produce different results for multiplication and rounding, this explains the divergence observed in the fake quantization experiment. Although not directly tested, it is highly likely that re-quantization in INT8 convolutions experiences similar inconsistencies, as it relies on multiplication and rounding internally.

\subsubsection{INT8 vs FP16}

Although the current state of INT8 quantization cannot achieve identical results across platforms---primarily due to non-standard implementations of rounding and multiplication in quantize/dequantize and re-quantization operations, this is not a fundamental limitation but rather a question of standardization. For example, re-quantization in INT8 convolutions could be implemented using integer-only arithmetic, provided that scales are represented as integers, eliminating the need for floating-point calculations. Even when floating-point quantization scales are used, cross-device reproducibility is possible if calculations adhere to the IEEE 754 standard, apply the same rounding mode, and maintain a consistent operation order. Such standardization, however, requires commitment from hardware vendors and will take time to achieve.

An alternative is to use FP16 convolutions. We performed a similar analysis as for INT8 convolutions to measure divergence across different devices, runtimes, and vendors. The results are shown in Table \ref{tab:fp16_conv}. Compared to the INT8 quantized results, FP16 convolutions are implemented consistently across all devices, except for CPU calculations on Apple via Core ML and Qualcomm via ORT. Among NPUs, the main outlier is Apple; however, the error values remain small, especially when compared to INT8 quantized convolutions. Across all other combinations of device, runtime, and vendor, the divergence is minimal.

\subsection{Calibration Information}
We replicate the experiment in Table 2 from \cite{tian2023towards}, using the DCVC-RT model.
The results are shown in Table~\ref{tab:cit-failures}.
A single video of 96 frames can be successfully coded across devices by adding a small amount of calibration information. However, this immediately breaks down as soon as we switch to FP16, the format necessary for NPU acceleration. 

Under the standard model of floating point arithmetic \footnote{For example presented \href{https://nhigham.com/wp-content/uploads/2021/04/high21m.pdf}{here}}, the upper bound of the absolute error from applying $n$ sequential floating point operations (such as a neural network) is bounded by:

\[ |y - \hat{y}| \leq \gamma_n |y|, \quad \text{where} \quad \gamma_n = \frac{nu}{1 - nu} \approx nu \]

Here $u$ is the unit roundoff and $n$ the number of operations. The measured maximum error between encoder and decoder float scale indices ($\ddot{I}$) is indeed approximately four orders of magnitude higher for FP16, as the theoretical model would predict ($u_{fp32} = 5.96 \times 10^{-8}$, $u_{fp16} = 4.88 \times 10^{-4})$.

\begin{table*}[t]
\centering
\caption{Calibration Information experiment using DCVC-RT, similar to Table 2 in \cite{tian2023towards}. The table shows the number of successfully decoded frames (with maximum absolute error between raw indices on coding first frame) across different platforms. Calibration Information does not resolve index mismatches in FP16.}
\label{tab:cit-failures}
\setlength{\tabcolsep}{6pt}
\begin{tabular*}{\textwidth}{@{\extracolsep{\fill}}lcccccc}
\toprule
& \multicolumn{3}{c}{Compressed on CPU 1 (fp32)} & \multicolumn{3}{c}{Compressed on NPU 1 (fp16)} \\
\cmidrule(lr){2-4}\cmidrule(lr){5-7}
& CPU 1 & CPU 2 & CPU 3 & NPU 1 & GPU 1 & CPU 2 \\
\midrule
\multicolumn{7}{l}{\textit{Test data:} 96 frames of VCD  0380a333cb0fef69001fb4260e2a705f video (640$\times$360)} \\
\midrule
w/o CIT & 96 (0) & \textcolor{red}{65} (4e-5) & \textcolor{red}{71} (6e-5) & 96 (0) & \textcolor{red}{0} (4e-1) & \textcolor{red}{0} (6e-1) \\
w CIT ($\epsilon = 10^{-4}$) & 96 (0) & 96 (4e-5) & 96 (6e-5) & 96 (0) & \textcolor{red}{0} (4e-1) & \textcolor{red}{0} (6e-1) \\
w CIT ($\epsilon = 10^{-1}$) & 96 (0) & 96 (4e-5) & 96 (6e-5) & 96 (0) & \textcolor{red}{0} (4e-1) & \textcolor{red}{0} (6e-1) \\
\bottomrule
\end{tabular*}

\medskip
\footnotesize
CPU 1: Snapdragon(R) X 12-core X1E80100 \quad
CPU 2: Apple M3 Pro CPU \quad
CPU 3: Intel(R) Xeon(R) Platinum 8370C \quad
NPU 1: Apple M3 Pro Neural Engine \quad
GPU 1: Apple M3 Pro GPU
\end{table*}

\subsection{Activation Functions}
\label{sec:act_func_divergence}

\begin{figure*}[t]
  \centering
  \includegraphics[width=\linewidth]{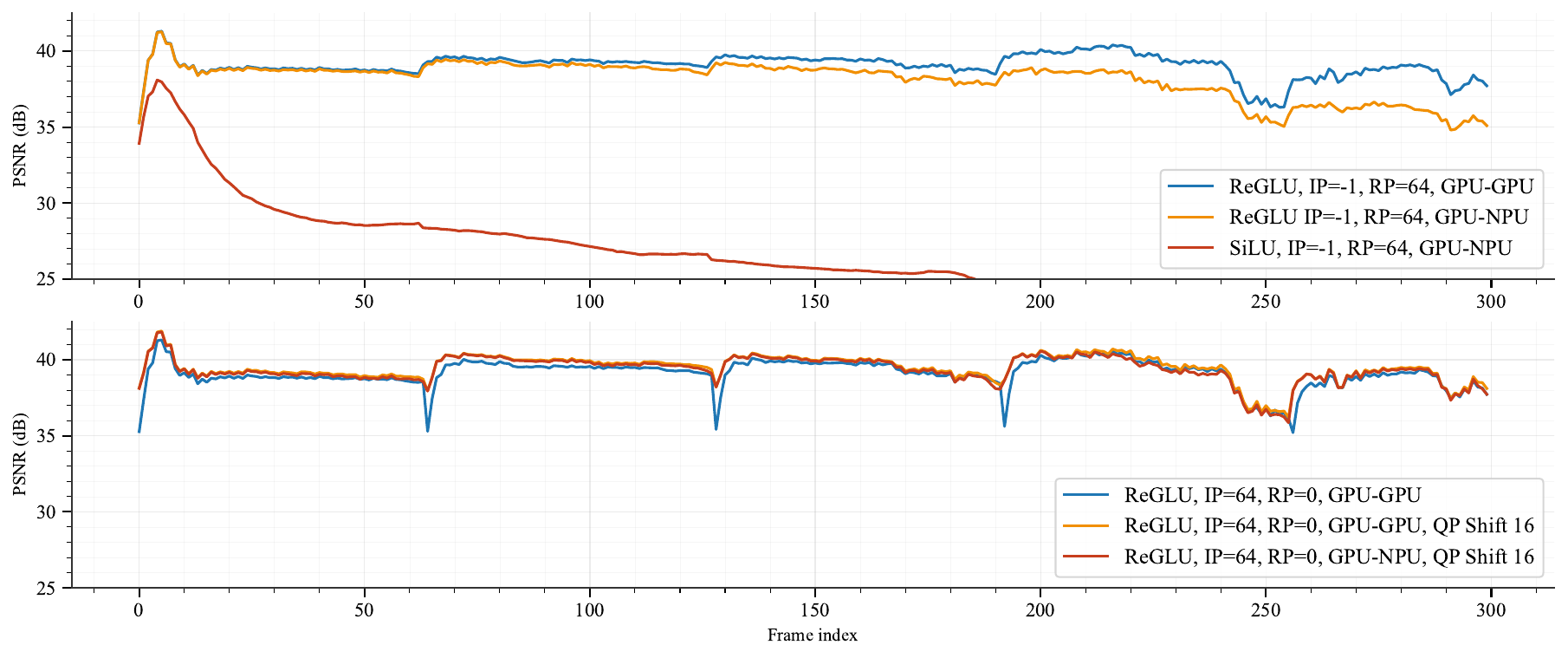}
  \caption{Example of cross-platform PSNR drift on a VCD-TH test sequence evaluated on Apple M3. The top row shows that a model using SiLU activations has rapid divergence, whereas ReGLU activations maintain stable predictions for substantially longer. To further suppress long-term drift, periodic intra frames are required. In our setting, the same P-frame model is reused for I-frames, which introduces an initial PSNR drop and can produce visible glitches. Increasing the QP of the independent frame reduces these artifacts.}
  \label{fig:psnr_over_time}
\end{figure*}

While some activations give better performance (ablation in Table \ref{tab:act_ablation}), they increase divergence between the encoder and decoder.
Figure \ref{fig:act_error} compares multiple functions on the Apple M3 NPU and CPU. LeakyReLU and ReLU are error-free, while other activations have significant differences.
Figure \ref{fig:silu_cpu_npu} presents a zoomed-in view for SiLU, showing that piece-wise approximation is used when running the activation on the NPU.
Figure \ref{fig:psnr_over_time} illustrates how such numerical differences accumulate over time, causing PSNR drift on a VCD-TH test sequence.

\begin{figure}[t]
  \centering
  \includegraphics[width=0.9\linewidth]{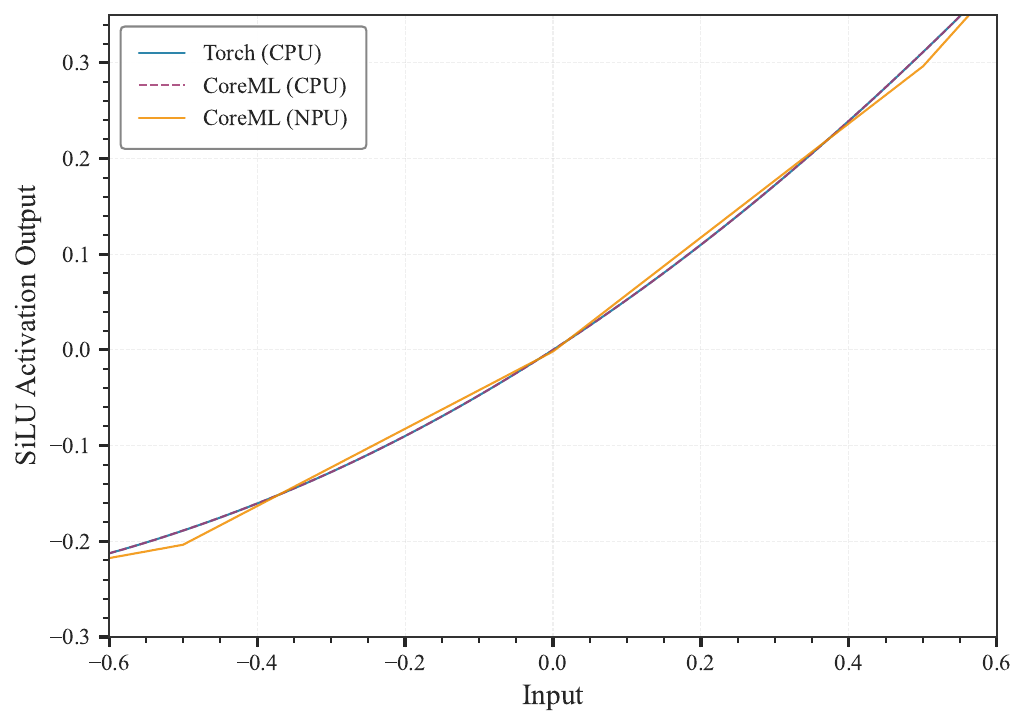}
  \caption{Numeric differences of SiLU activation function on Apple. A clear piece-wise approximation is apparent for the NPU.}
  \label{fig:silu_cpu_npu}
\end{figure}

\begin{figure*}[t]
  \centering
  \includegraphics[width=\linewidth]{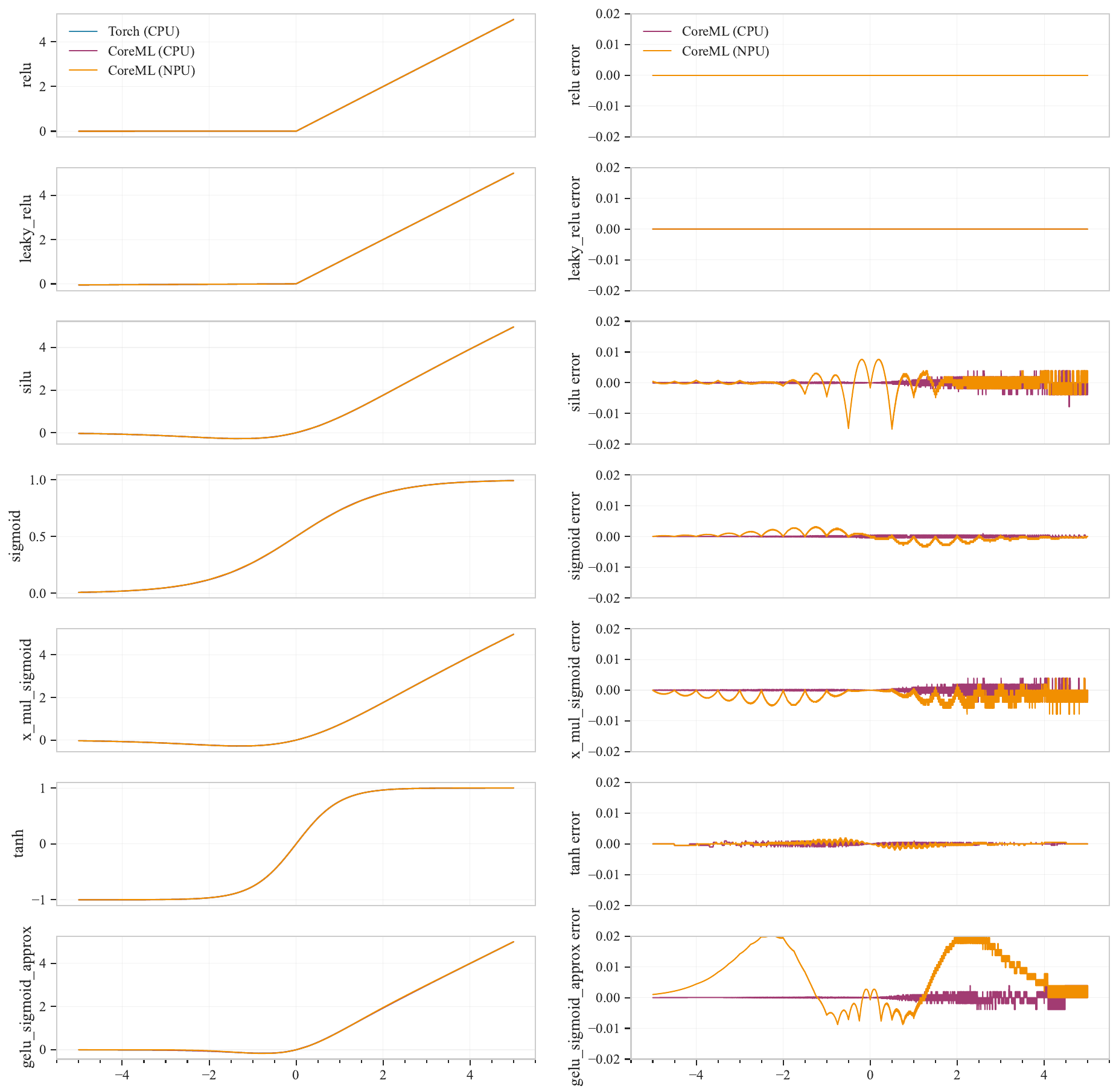}
  \caption{Activation function output divergence between Torch CPU implementation (reference) and CoreML CPU and NPU implementations run on Apple M3. LeakyReLU and ReLU show no error, while other activations show significant divergence.}
  \label{fig:act_error}
\end{figure*}

\begin{figure*}[t]
  \centering
  \includegraphics[width=\linewidth]{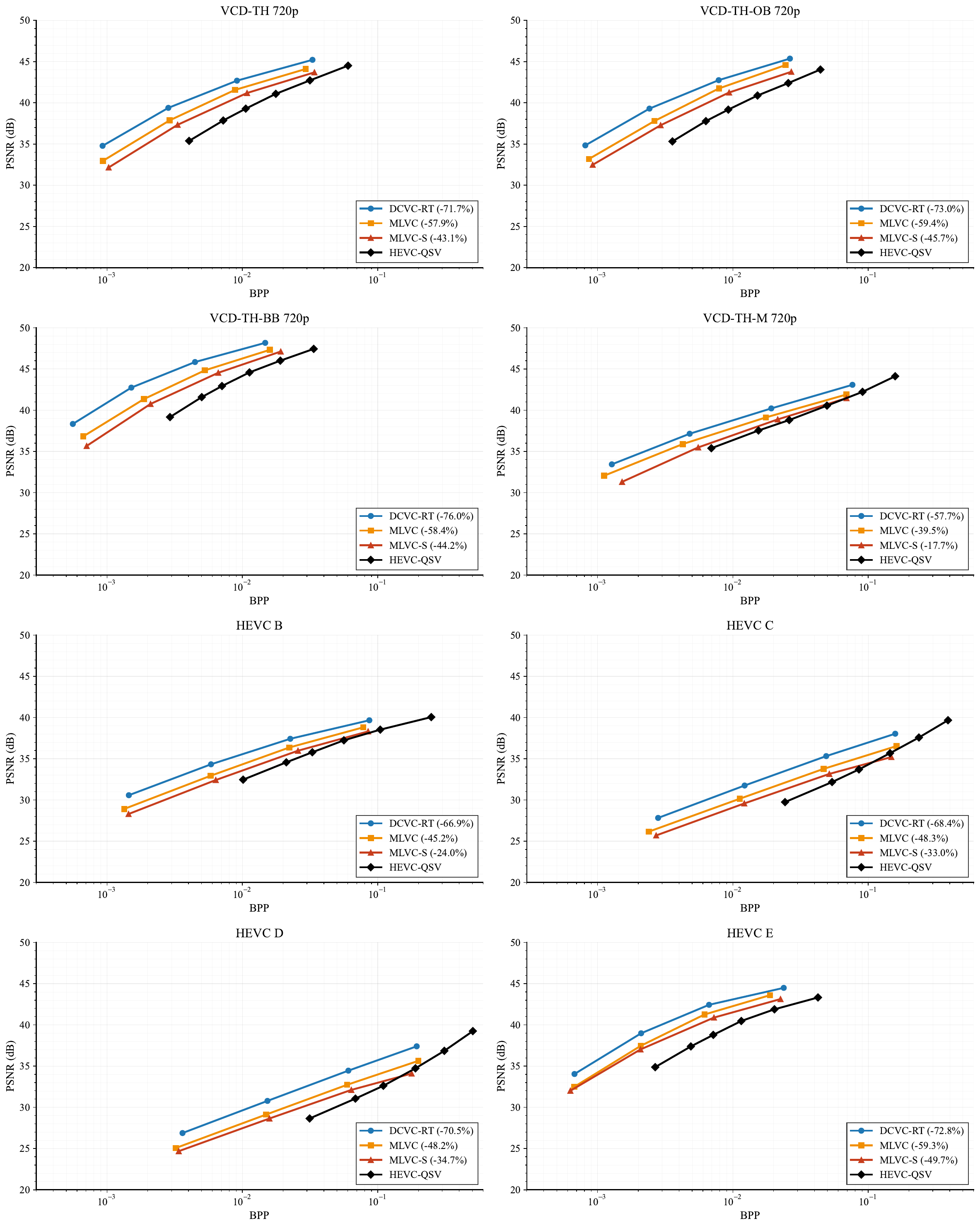}
  \caption{Rate-distortion curves for VCD-720p and HEVC datasets. The curves correspond to the first (DCVC-RT) and bottom two rows (MLVC, MLVC-S) in the ablation study Table \ref{tab:ablation}.}
  \label{fig:ablation_rd_curves}
\end{figure*}

\fi

\end{document}